\documentclass[journal]{IEEEtran}
\usepackage[T1]{fontenc}
\usepackage[latin9]{inputenc}
\usepackage{color}
\usepackage{array}
\usepackage{float}
\usepackage{multirow}
\usepackage{amsmath}
\usepackage{amssymb}
\usepackage{graphicx}
\usepackage{setspace}
\usepackage[bookmarks=true,bookmarksnumbered=true,bookmarksopen=true,bookmarksopenlevel=1,
 breaklinks=false,pdfborder={0 0 0},pdfborderstyle={},backref=false,colorlinks=false]
 {hyperref}
\hypersetup{pdftitle={Your Title},
 pdfauthor={Your Name},
 pdfpagelayout=OneColumn, pdfnewwindow=true, pdfstartview=XYZ, plainpages=false}

\makeatletter

\providecommand{\tabularnewline}{\\}
\floatstyle{ruled}
\newfloat{algorithm}{tbp}{loa}
\providecommand{\algorithmname}{Algorithm}
\floatname{algorithm}{\protect\algorithmname}

\usepackage[caption=false,font=footnotesize]{subfig}
\usepackage{algorithmic}
\usepackage{cite}

\usepackage{caption}

\ifdefined\showcaptionsetup
 \PassOptionsToPackage{caption=false}{subfig}
\fi
\usepackage{subfig}
\makeatother

\begin{document}
\title{Exploiting Dynamic Sparsity for Near-Field Spatial Non-Stationary
XL-MIMO Channel Tracking}
\author{Wenkang~Xu,~An Liu, \IEEEmembership{Senior Member,~IEEE,} Min-jian
Zhao{\normalsize , }\IEEEmembership{Member,~IEEE,}{\normalsize{} Giuseppe
Caire, }\IEEEmembership{Fellow,~IEEE}, and Yik-Chung Wu, \IEEEmembership{Senior Member,~IEEE}{\normalsize{}
\thanks{Wenkang Xu, An Liu, and Min-jian Zhao are with the College of Information
Science and Electronic Engineering, Zhejiang University, Hangzhou
310027, China (email: anliu@zju.edu.cn).

G. Caire is with the Department of Telecommunication Systems, Technical
University of Berlin, 10587 Berlin, Germany (e-mail: caire@tu-berlin.de).}\thanks{Yik-Chung Wu is with the Department of Electrical and Electronic Engineering,
The University of Hong Kong, Hong Kong (e-mail: ycwu@eee.hku.hk).}\vspace{-2mm}
}}
\maketitle
\begin{abstract}
This work considers a spatial non-stationary channel tracking problem
in broadband extremely large-scale multiple-input-multiple-output
(XL-MIMO) systems. In the case of spatial non-stationary, each scatterer
has a certain visibility region (VR) over antennas and power change
may occur among visible antennas. Concentrating on the temporal correlation
of XL-MIMO channels, we design a three-layer Markov prior model and
hierarchical two-dimensional (2D) Markov model to exploit the dynamic
sparsity of sparse channel vectors and VRs, respectively. Then, we
formulate the channel tracking problem as a bilinear measurement process,
and a novel dynamic alternating maximum a posteriori (DA-MAP) framework
is developed to solve the problem. The DA-MAP contains four basic
modules: channel estimation module, VR detection module, grid update
module, and temporal correlated module. Specifically, the first module
is an inverse-free variational Bayesian inference (IF-VBI) estimator
that avoids computational intensive matrix inverse each iteration;
the second module is a turbo compressive sensing (Turbo-CS) algorithm
that only needs small-scale matrix operations in a parallel fashion;
the third module refines the polar-delay domain grid; and the fourth
module can process the temporal prior information to ensure high-efficiency
channel tracking. Simulations show that the proposed method can achieve
a significant channel tracking performance while achieving low computational
overhead.
\end{abstract}


\section{Introduction}

Extremely large-scale multiple-input-multiple-output (XL-MIMO) is
a promising technology with many applications for future 6G communications
\cite{Rappaport_survey}. With an even larger number of antennas at
the base station (BS) compared to traditional massive MIMO systems,
XL-MIMO can offer higher data rates, improved spectral efficiency,
and enhanced system reliability \cite{Lu_NF_Tutorial}. To reap the
benefits of XL-MIMO, it is essential to track channel state information
(CSI) accurately over a time-varying wireless channel \cite{Ziniel_AMP_Tracking,Lian_FDD_Tracking,LiuGY_Angular_Tracking,WanYB_5G_Tracking}.

There are a number of characteristics of XL-MIMO channels that distinguish
it from conventional MIMO channels \cite{Liu_NF_Tutorial,Cui_polar_grid,Carvalho_NSN,Yuan_NSN}.
First of all, the effect of near-field propagation is obvious. Due
to the large antenna aperture of XL-MIMO, the near-field region (Rayleigh
region) of the array is also large \cite{Liu_NF_Tutorial}. The array
experiences a spherical wave within the Rayleigh region, and the resulting
steering vector accounts for both angle and distance parameters of
last-hop scatterers \cite{Cui_polar_grid}. Secondly, the spatial
non-stationary effect is likely to exist. Spatial non-stationary means
that each scatterer can only see a subset of antennas, i.e., it has
a certain visibility region (VR) over the antennas \cite{Carvalho_NSN}.
Furthermore, the measurements in \cite{Yuan_NSN} revealed that power
change among visible antenna elements may occur in some channel paths.
In this case, VR needs to be modeled as a continuous variable instead
of a 0/1 discrete variable, which makes the VR issue more challenging.
Thirdly, the XL-MIMO channel varies quickly over time. Fortunately,
there is a temporal correlation among XL-MIMO channels. Specifically,
although the channel varies quickly over time, the scattering environment,
which consists of the angle, distance, and VR of scatterers, changes
quite slowly in adjacent frames. By exploiting this temporal correlation,
the channel tracking performance is expected to be enhanced significantly.
Some related works on how to estimate and track spatial non-stationary
channels are summarized below.

\textbf{Spatial non-stationary channel estimation: }There are mainly
three types of VR modeling methods, and different VR modeling methods
lead to different algorithm designs for channel estimation. The first
one is called the sub-array-level 0/1 modeling. In this model, the
large-scale array are divided into a number of small-scale sub-arrays,
and each sub-channel corresponding to each sub-array is assumed to
be spatial stationary. Based on the sub-array grouping, binary vectors
are used to model the VR of scatterers and sub-arrays. Such a modeling
method was firstly introduced in \cite{HanYu_VR}, and a sub-array-wise
orthogonal matching pursuit (OMP) algorithm was designed for channel
estimation. Inspired by this, many sub-array-wise methods were developed
for addressing the spatial non-stationary issue \cite{HanYu_VR2,Iimori_VR,Chen_NS_CE}.
Generally speaking, the sub-array-wise methods first estimate each
spatial stationary sub-channel independently, and then combine the
obtained sub-channels into the whole channel. Apparently, they work
poorly in the low SNR regions since the correlation among sub-channels
is ignored. Moreover, the spatial stationary of each sub-array cannot
be ensured in practice. To overcome this drawback, the antenna-level
0/1 modeling for VR was used in \cite{Vincent_sparse_modeling,Tang_sparse_modeling}.
This model considers the VR of each antenna without sub-array grouping,
and the visible antennas are concentrated on a few clusters. The work
in \cite{Vincent_sparse_modeling} used a Markov chain to capture
the clustered sparsity of the VR and proposed a turbo orthogonal approximate
message passing (Turbo-OAMP) algorithm. However, such an algorithm
directly processes the received signal in the space domain, which
cannot exploit the polar-domain sparsity of XL-MIMO channels \cite{Cui_polar_grid}.
In \cite{Tang_sparse_modeling}, the authors developed a two-stage
method to exploit the sparsity of channels. But it was only confined
to the single-path scenario. For more general multi-path scenario,
\cite{Xu_alternating_MAP} proposed a novel alternating maximum a
posteriori (MAP) framework to achieve joint VR detection and channel
estimation. Recently, power change among visible antennas was observed
in \cite{Yuan_NSN}. A more accurate VR modeling method was proposed,
in which the VR was no longer based on the 0/1 assumption but treated
as a continuous variable. Motivated by this, \cite{TonyQS_XL-RIS}
developed a robust fast sparse Bayesian learning (RFSBL) algorithm
to estimate the continuous VRs.

\textbf{Multi-frame channel tracking: }The above works focus on channel
estimation for each frame separately, while the temporal correlation
among multiple frames is not considered. To the best of our knowledge,
there is still a lack of research on spatial non-stationary channel
tracking. Nevertheless, some early attempts at massive MIMO channel
tracking inspire us a lot. In particular, \cite{Ziniel_AMP_Tracking}
formulated the multi-frame tracking problem as a dynamic compressive
sensing (CS) problem, and an approximate message passing (AMP) algorithm
is used to solve the problem. The work in \cite{Lian_FDD_Tracking}
used a Markov chain to model the temporal correlation of channels
and designed a Turbo-OAMP algorithm to track the dynamic channels
with the Markov prior. \cite{LiuGY_Angular_Tracking,WanYB_5G_Tracking}
considered practical system imperfections and tracked the dynamic
channels and system imperfections jointly. Although these methods
have elaborated on how to exploit the dynamic sparsity of spatial
stationary channels, they cannot be directly applied to the case of
spatial non-stationary channels. The reasons are two folds: 1) how
to exploit the temporal correlation of VRs is still known; 2) these
algorithms are used to deal with the linear measurement process, while
the considered spatial non-stationary channel tracking problem is
a bilinear measurement process.

In this work, we consider a non-stationary channel tracking problem
in a broadband multi-carrier XL-MIMO system under a hybrid beamforming
(HBF) architecture. Based on the literature review, there are some
challenges in the considered problem. Firstly, it is important to
design a proper prior model to capture the temporal correlation of
both sparse channel vectors and continuous VRs. Secondly, the considered
problem is a bilinear measurement process, and thus a new algorithmic
framework needs to be developed to solve the problem. Thirdly, due
to the large number of antennas and sub-carriers, the dimension of
XL-MIMO channels is extremely high. Therefore, low computational overhead
is also essential while acquiring CSI accurately. This paper aims
to address these challenges and the main contributions are summarized
as follows.
\begin{itemize}
\item \textbf{Encoded sub-array HBF architecture:} The received signals
of antennas are mixed together due to HBF, which makes it difficult
to decouple the individual received signal at each antenna for further
VR detection. Inspired by \cite{Chen_NS_CE}, a more generalized signal
extraction scheme is developed based on an encoded sub-array HBF architecture.
\item \textbf{Prior design for the dynamic sparsity:} A three-layer Markov
prior model is designed to capture the dynamic sparsity of the polar-delay
domain channel vectors in multiple frames. Besides, we introduce a
hierarchical two-dimensional (2D) Markov model to exploit the clustered
sparsity of VRs over antennas and correlation over time. 
\item \textbf{Dynamic alternating MAP (DA-MAP): }A novel DA-MAP framework
is proposed to deal with the bilinear measurement process. The DA-MAP
framework consists of four basic modules: channel estimation module,
VR detection module, grid update module, and temporal correlated module.
Specifically, the first module is used to estimate the polar-delay
domain channel vectors, while the second module is employed to recover
continuous VRs. Furthermore, the third module refines the polar-delay
domain grid, and the fourth module handles the temporal prior information
passed from the previous frame to the current frame. For each frame,
the channel estimation module, VR detection module, and grid update
module work alternatively to perform high-accuracy spatial non-stationary
channel estimation. Among multiple frames, the temporal correlated
module exploits the temporal correlation to achieve robust channel
tracking.
\item \textbf{Low-complexity designs:} Low-complexity designs are introduced
to reduce the computational complexity of the DA-MAP framework. For
the channel estimation module, we use an inverse-free variational
Bayesian inference (IF-VBI) estimator \cite{Xu_alternating_MAP,Xu_Turbo-IFVBI}
that avoids the computational intensive matrix inverse operation.
Besides, the turbo compressive sensing (Turbo-CS) algorithm \cite{LiuAn_CE_Turbo_CS}
is designed as the VR detection module, which only needs small-scale
matrix operations in a parallel fashion by resorting to polar-delay
domain filtering. Furthermore, exploiting information from the previous
frame as the prior for the next one, the iteration number of the DA-MAP
is reduced significantly during channel tracking. As a result, the
DA-MAP obtains an excellent channel tracking performance while achieving
low computational overhead. 
\end{itemize}

The rest of the paper proceeds as follows. Section II presents the
channel model and encoded sub-array HBF architecture. Section III
formulates the considered problem into a bilinear measurement process.
Section IV elaborates on the DA-MAP framework. The simulations results
are shown in Section V. Finally, the paper is concluded in Section
VI.

\textit{Notations:} Lowercase and uppercase bold letterers denote
vectors and matrices, respectively. Let $\left(\cdot\right)^{-1}$,
$\left(\cdot\right)^{T}$, $\left(\cdot\right)^{H}$, $\left\langle \cdot\right\rangle $,
$\left\Vert \cdot\right\Vert $, $\textrm{vec}\left(\cdot\right)$,
and $\textrm{diag}\left(\cdot\right)$ represent the inverse, transpose,
conjugate transpose, expectation, $\ell_{2}\textrm{-norm}$, vectorization,
and diagonalization operations, respectively. $\otimes$ is the Kronecker
product operator and $\odot$ means the Hadamard product operator.
$\Re\left\{ \cdot\right\} $ and $\Im\left\{ \cdot\right\} $ denote
the real and imaginary part of the complex argument, respectively.
$\mathbf{I}_{N}$ is the $N\times N$ dimensional identity matrix
and $\mathbf{1}_{M\times N}$ is the $M\times N$ dimensional all-one
matrix. For a set $\mathbf{\Omega}$ with its cardinal number denoted
by $\left|\mathbf{\Omega}\right|$, $\boldsymbol{x}\triangleq\left[x_{n}\right]_{n\in\mathbf{\Omega}}\in\mathbb{C}^{\left|\mathbf{\Omega}\right|\times1}$
is a vector composed of elements indexed by $\mathbf{\Omega}$. $\mathcal{CN}\left(\boldsymbol{x};\boldsymbol{\mu},\mathbf{\Sigma}\right)$
denotes the complex Gaussian distribution with mean $\boldsymbol{\mu}$
and covariance $\mathbf{\Sigma}$. $\textrm{Ga}\left(x;a,b\right)$
denotes the Gamma distribution with shape parameter $a$ and rate
parameter $b$.

\section{System Model}

\subsection{Broadband XL-MIMO System under Hybrid Beamforming}

\begin{figure}[t]
\centering{}\includegraphics[width=90mm]{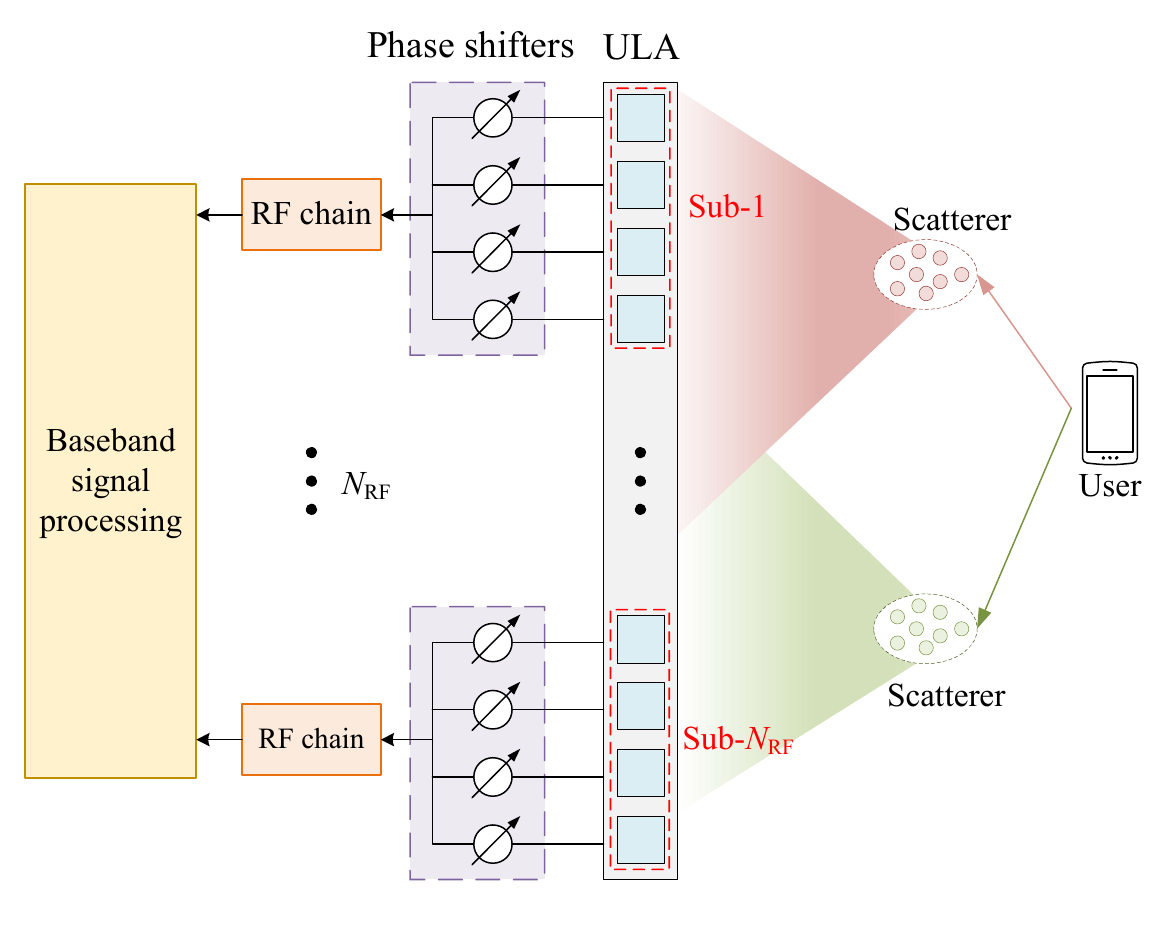}\caption{\label{fig:system_model}Illustration of the spatial non-stationary
XL-MIMO channel and the sub-array HBF architecture.}
\end{figure}

Consider an uplink time division duplexing (TDD) based broadband XL-MIMO
system, where a BS serves single-antenna users, as illustrated in
Fig. \ref{fig:system_model}. The BS is equipped with a half-wavelength
uniform linear array (ULA) of $M$ antennas and a sub-array HBF architecture
with $N_{\textrm{RF}}\ll M$ RF chains.\footnote{The proposed method can be easily extended to the uniform planar array
(UPA) system by replacing the steering vector in (\ref{eq:a(theta,r)_new})
\cite{Xu_alternating_MAP}. } Assume that $N_{\textrm{RF}}$ is divisible by $M$ and partition
the ULA into $N_{\textrm{RF}}$ sub-arrays uniformly, with each sub-array
containing $M_{\textrm{sub}}=\frac{M}{N_{\textrm{RF}}}$ antennas.
All the antennas in the same sub-array is only connected with one
RF chain, which can effectively reduce hardware complexity \cite{Lu_CE_UL}.
Each single-antenna user is allocated with a subset of $N$ subcarriers
for channel estimation. Since different users are assigned with orthogonal
frequency resources during channel estimation \cite{Fundamentals},
we shall focus on a certain user. The center frequency is $f_{c}$,
the subcarrier interval is $f_{0}$, and the wavelength is $\lambda_{c}=\frac{c}{f_{c}}$,
where $c$ is the speed of light.

In the $t\textrm{-th}$ frame, the user transmits a pilot sequence
of length $P$ to the BS for channel estimation. Then, the received
signal of the $p\textrm{-th}$ pilot symbol on the $n\textrm{-th}$
subcarrier, denoted by $\boldsymbol{y}_{p}^{\left(t\right)}\left[n\right]\in\mathbb{C}^{N_{\textrm{RF}}\times1}$,
is expressed as
\begin{equation}
\boldsymbol{y}_{p}^{\left(t\right)}\left[n\right]=\mathbf{W}_{p}^{\left(t\right)}\boldsymbol{h}^{\left(t\right)}\left[n\right]v_{p}^{\left(t\right)}\left[n\right]+\mathbf{W}_{p}^{\left(t\right)}\boldsymbol{z}_{p}^{\left(t\right)}\left[n\right],\label{eq:y(t)p=00005Bn=00005D}
\end{equation}
where $\boldsymbol{h}^{\left(t\right)}\left[n\right]\in\mathbb{C}^{M\times1}$
is the channel response vector that is assumed to stay unchanged within
each frame, $v_{p}^{\left(t\right)}\left[n\right]\in\mathbb{C}$ is
the pilot symbol with unit power, and $\boldsymbol{z}_{p}^{\left(t\right)}\left[n\right]\in\mathbb{C}^{M\times1}$
is the additive white Gaussian noise (AWGN) with variance $1/\gamma_{z}^{\left(t\right)}$.
The $\mathbf{W}_{p}^{\left(t\right)}\in\mathbb{C}^{N_{\textrm{RF}}\times M}$
in (\ref{eq:y(t)p=00005Bn=00005D}) is the constant amplitude analog
phase shifter matrix, which is defined as
\begin{equation}
\mathbf{W}_{p}^{\left(t\right)}\triangleq\textrm{BlockDiag}\left(\left(\boldsymbol{w}_{p,1}^{\left(t\right)}\right)^{T},\ldots,\left(\boldsymbol{w}_{p,N_{\textrm{RF}}}^{\left(t\right)}\right)^{T}\right).\label{eq:BLOCKW}
\end{equation}
where $\textrm{BlockDiag}\left(\cdot\right)$ denotes the block diagonal
operator, and the analog phase shifter vector for the $k\textrm{-th}$
sub-array is $\boldsymbol{w}_{p,k}^{\left(t\right)}\triangleq\left[e^{j\varphi_{p,k,1}^{\left(t\right)}},\ldots,e^{j\varphi_{p,k,M_{\textrm{sub}}}^{\left(t\right)}}\right]^{T}\in\mathbb{C}^{M_{\textrm{sub}}\times1}$
with $\varphi_{p,k,m}^{\left(t\right)}\in\left[0,2\pi\right]$ for
$m=1,\ldots,M_{\textrm{sub}}$. 

Due to the near-field propagation and spatial non-stationary, the
XL-MIMO channel $\boldsymbol{h}^{\left(t\right)}\left[n\right]$ is
quite different from the traditional massive MIMO channel. The modeling
of $\boldsymbol{h}^{\left(t\right)}\left[n\right]$ is discussed next.

\subsection{Spatial Non-stationary Channel Model}

In a high-frequency XL-MIMO system, the Rayleigh distance is hundreds
or even thousands of meters \cite{Liu_NF_Tutorial,Rappaport_survey}.
In this case, near-field propagation cannot be ignored, and the steering
vector is related to both the angle and distance parameters of scatterers.
Without loss of generality, we set the center of the ULA as the origin
of the coordinate system. Let $\delta_{m}=m-\frac{M+1}{2}$ represent
the relative index of the $m\textrm{-th}$ antenna. Then the coordinates
of the $m\textrm{-th}$ antenna can be expressed as $\boldsymbol{p}_{m}=\left[\delta_{m}d,0\right]^{T}$,
where $d=\frac{\lambda_{c}}{2}$ is the antenna spacing. 

Consider a scatterer with angle of arrival (AoA) $\theta$ and distance
$r$, whose coordinates is $\boldsymbol{p}=\left[r\cos\theta,r\sin\theta\right]^{T}$.
The distance between the scatterer and the $m\textrm{-th}$ antenna
can be derived as
\begin{equation}
r_{m}=\left\Vert \boldsymbol{p}-\boldsymbol{p}_{m}\right\Vert =\sqrt{r^{2}+\delta_{m}^{2}d^{2}-2r\delta_{m}d\cos\theta}.
\end{equation}
Based on the uniform spherical wave (USW) model \cite{Starer_USW1},
the steering vector at the BS is given by
\begin{equation}
\boldsymbol{a}\left(\theta,r\right)=\frac{1}{\sqrt{M}}\left[e^{-j\frac{2\pi}{\lambda_{c}}\left(r_{1}-r\right)},\ldots,e^{-j\frac{2\pi}{\lambda_{c}}\left(r_{M}-r\right)}\right]^{T}.\label{eq:a(theta,r)_old}
\end{equation}
Since the form in (\ref{eq:a(theta,r)_old}) is quite complicated,
we introduce the Fresnel approximation to approximate the distance
$r_{m}$ \cite{Selvan_Fresnel_distance},
\begin{equation}
r_{m}\approx r-\delta_{m}d\vartheta+\frac{\delta_{m}^{2}d^{2}}{2r}\left(1-\vartheta^{2}\right),\label{eq:rm_new}
\end{equation}
where $\vartheta\triangleq\cos\theta$. The above Fresnel approximation
has been verified to be accurate enough and widely used in near-field
channel modeling \cite{Liu_NF_Tutorial,Cui_polar_grid,Pan_THz_NFCE}.
Submitting (\ref{eq:rm_new}) into (\ref{eq:a(theta,r)_old}), the
simplified steering vector based on the Fresnel approximation is obtained
as
\begin{equation}
\left[\boldsymbol{a}\left(\vartheta,r\right)\right]_{m}=\frac{1}{\sqrt{M}}e^{-j\frac{2\pi}{\lambda_{c}}\left(-\delta_{m}d\vartheta+\frac{\delta_{m}^{2}d^{2}}{2r}\left(1-\vartheta^{2}\right)\right)},\label{eq:a(theta,r)_new}
\end{equation}
where $\left[\boldsymbol{a}\left(\vartheta,r\right)\right]_{m}$ is
the $m\textrm{-th}$ element of $\boldsymbol{a}\left(\vartheta,r\right)$.

Besides, the spatial non-stationary effect means that each scatterer
has a certain VR over antennas. Let $L^{\left(t\right)}$ denote the
number of channel paths in $\boldsymbol{h}^{\left(t\right)}\left[n\right]$.
Define $\boldsymbol{u}_{l}^{\left(t\right)}\triangleq\Bigl[u_{l,1}^{\left(t\right)},\ldots,u_{l,M}^{\left(t\right)}\Bigr]^{T}\in\mathbb{C}^{M\times1}$
as the VR vector of scatterer $l$ in the $t\textrm{-th}$ frame,
where $u_{l,m}^{\left(t\right)}\geq0$ represents the received power
of the $l\textrm{-th}$ path on the $m\textrm{-th}$ antenna with
$u_{l,m}^{\left(t\right)}=0$ indicating the $m\textrm{-th}$ antenna
is invisible to scatterer $l$ \cite{Yuan_NSN}. Consequently, the
channel vector is modeled as
\begin{equation}
\boldsymbol{h}^{\left(t\right)}\left[n\right]=\sum_{l=1}^{L^{\left(t\right)}}x_{l}^{\left(t\right)}e^{-j2\pi nf_{0}\tau_{l}^{\left(t\right)}}\boldsymbol{a}\left(\vartheta_{l}^{\left(t\right)},r_{l}^{\left(t\right)}\right)\odot\boldsymbol{u}_{l}^{\left(t\right)},\label{eq:h(t)=00005Bn=00005D}
\end{equation}
where $x_{l}^{\left(t\right)}$, $\tau_{l}^{\left(t\right)}$, $\vartheta_{l}^{\left(t\right)}$,
and $r_{l}^{\left(t\right)}$ represent the complex channel gain,
delay, angle, and distance of the $l\textrm{-th}$ path.

\subsection{Encoded Sub-array HBF Architecture}

From (\ref{eq:y(t)p=00005Bn=00005D}), we notice that the received
signal of each RF chain is a mixture of the received signal of antennas
in its associated sub-array. In \cite{Chen_NS_CE}, a signal extraction
scheme is developed when a fully connected HBF architecture is used
for $\mathbf{W}_{p}^{\left(t\right)}$. However, the number of antennas
is confined to $M=2^{i},i\in\mathbb{N}+$ and the number of pilots
is confined to $P=N_{P}M,N_{P}\in\mathbb{N}+$. Moreover, it needs
at least $P=M$ pilot observations to decouple the signals from each
antenna, which leads to a large pilot overhead and is unacceptable
for practical XL-MIMO systems. To overcome these drawbacks, we make
use of the block diagonal structure in (\ref{eq:BLOCKW}) and propose
a more generalized signal extraction scheme based on an encoded sub-array
HBF architecture. 

In particular, the received signal of the $k\textrm{-th}$ RF chain
can be expressed as
\begin{equation}
y_{p,k}^{\left(t\right)}\left[n\right]=\left(\boldsymbol{w}_{p,k}^{\left(t\right)}\right)^{T}\left(\boldsymbol{h}_{\textrm{sub-}k}^{\left(t\right)}\left[n\right]v_{p}^{\left(t\right)}\left[n\right]+\boldsymbol{z}_{p,\textrm{sub-}k}^{\left(t\right)}\left[n\right]\right),\label{eq:y_pk(t)=00005Bn=00005D}
\end{equation}
where $\boldsymbol{h}_{\textrm{sub-}k}^{\left(t\right)}\left[n\right]\triangleq\left[h_{m}^{\left(t\right)}\left[n\right]\right]_{m\in\mathbf{\Psi}_{k}}$
and $\boldsymbol{z}_{p,\textrm{sub-}k}^{\left(t\right)}\left[n\right]\triangleq\left[z_{p,m}^{\left(t\right)}\left[n\right]\right]_{m\in\mathbf{\Psi}_{k}}$
denote the channel vector and noise vector corresponding to the $k\textrm{-th}$
sub-array, respectively, and $\mathbf{\Psi}_{k}\triangleq\left\{ m+\left(k-1\right)M_{\textrm{sub}}\mid m=1,\ldots,M_{\textrm{sub}}\right\} $
is the index set of antennas in the $k\textrm{-th}$ sub-array. To
simplify the notation, we temporarily omit the indices of subcarriers
and frames. Then, (\ref{eq:y_pk(t)=00005Bn=00005D}) can be rewritten
into
\begin{equation}
y_{p,k}=\left(\boldsymbol{w}_{p,k}\right)^{T}\left(\boldsymbol{h}_{\textrm{sub-}k}v_{p}+\boldsymbol{z}_{p,\textrm{sub-}k}\right),\forall p.\label{eq:y_pk}
\end{equation}
The goal is to acquire the individual received signal at $M_{\textrm{sub}}$
antennas from $P$ linear mixture measurements of (\ref{eq:y_pk}),
where $P\geqslant M_{\textrm{sub}}$ is a necessary condition.

Focusing on the case of $P\geqslant M_{\textrm{sub}}$, we encode
the phase shifter vectors as
\begin{equation}
\left[\boldsymbol{w}_{1,k},\ldots,\boldsymbol{w}_{P,k}\right]^{T}=\mathbf{D}_{P},\label{eq:HFB_edcoding}
\end{equation}
where $\mathbf{D}_{P}\in\mathbb{C}^{P\times M_{\textrm{sub}}}$ consists
of the first $M_{\textrm{sub}}$ columns of the $P$-dimensional discrete
Fourier transform (DFT) matrix such that $\mathbf{D}_{P}^{H}\mathbf{D}_{P}=P\mathbf{I}_{M_{\textrm{sub}}}$.
Then, the received signal of each antenna can be decoupled as
\begin{align}
\tilde{\boldsymbol{y}}_{\textrm{sub-}k} & =\frac{1}{P}\mathbf{D}_{P}^{H}\left[\frac{y_{1,k}}{v_{1}},\ldots,\frac{y_{P,k}}{v_{P}}\right]^{T},\nonumber \\
 & =\boldsymbol{h}_{\textrm{sub-}k}+\tilde{\boldsymbol{z}}_{\textrm{sub-}k},\label{eq:HBF_decoding}
\end{align}
where $\tilde{\boldsymbol{z}}_{\textrm{sub-}k}\in\mathbb{C}^{M_{\textrm{sub}}\times1}$
is the equivalent noise vector with variance $\gamma^{\left(t\right)}=\frac{P}{M_{\textrm{sub}}\gamma_{z}^{\left(t\right)}}$. 

Stacking the decoupled signals of all RF chains into a single vector
gives
\begin{equation}
\tilde{\boldsymbol{y}}^{\left(t\right)}\left[n\right]=\boldsymbol{h}^{\left(t\right)}\left[n\right]+\tilde{\boldsymbol{z}}^{\left(t\right)}\left[n\right],\label{eq:y_tilde(t)=00005Bn=00005D}
\end{equation}
where
\begin{align*}
\tilde{\boldsymbol{y}}^{\left(t\right)}\left[n\right] & \triangleq\left[\left(\tilde{\boldsymbol{y}}_{\textrm{sub-}1}^{\left(t\right)}\left[n\right]\right)^{T},\ldots,\left(\tilde{\boldsymbol{y}}_{\textrm{sub-}N_{\textrm{RF}}}^{\left(t\right)}\left[n\right]\right)^{T}\right]^{T}\in\mathbb{C}^{M\times1},\\
\tilde{\boldsymbol{z}}^{\left(t\right)}\left[n\right] & \triangleq\left[\left(\tilde{\boldsymbol{z}}_{\textrm{sub-}1}^{\left(t\right)}\left[n\right]\right)^{T},\ldots,\left(\tilde{\boldsymbol{z}}_{\textrm{sub-}N_{\textrm{RF}}}^{\left(t\right)}\left[n\right]\right)^{T}\right]^{T}\in\mathbb{C}^{M\times1}.
\end{align*}
Further stacking $\tilde{\boldsymbol{y}}^{\left(t\right)}\left[n\right]$
for all $N$ subcarriers, we obtain
\begin{equation}
\tilde{\mathbf{Y}}^{\left(t\right)}=\mathbf{H}^{\left(t\right)}+\tilde{\mathbf{Z}}^{\left(t\right)},\label{eq:Y_tilde(t)}
\end{equation}
where $\mathbf{H}^{\left(t\right)}\triangleq\left[\left(\boldsymbol{h}^{\left(t\right)}\left[1\right]\right)^{T};\ldots;\left(\boldsymbol{h}^{\left(t\right)}\left[N\right]\right)^{T}\right]\in\mathbb{C}^{N\times M}$
is the frequency-antenna domain channel matrix, and $\tilde{\mathbf{Y}}^{\left(t\right)}$
and $\tilde{\mathbf{Z}}^{\left(t\right)}$ are defined similarly.
Vectorization of $\tilde{\mathbf{Y}}^{\left(t\right)}$ yields
\begin{equation}
\tilde{\boldsymbol{y}}^{\left(t\right)}=\boldsymbol{h}^{\left(t\right)}+\tilde{\boldsymbol{z}}^{\left(t\right)},\label{eq:y_tilde(t)}
\end{equation}
where $\tilde{\boldsymbol{y}}^{\left(t\right)}=\textrm{vec}\left(\tilde{\mathbf{Y}}^{\left(t\right)}\right)$,
$\boldsymbol{h}^{\left(t\right)}=\textrm{vec}\left(\mathbf{H}^{\left(t\right)}\right)$,
and $\tilde{\boldsymbol{z}}^{\left(t\right)}=\textrm{vec}\left(\tilde{\mathbf{Z}}^{\left(t\right)}\right)$.

\section{Prior Design for the Dynamic Sparsity}

\subsection{Polar-delay Domain Sparse Representation}

Due to the small number of paths constituting the channel, grid-based
sparse representation of wireless channels is commonly used \cite{OGSBI}.
Recently, the work in \cite{Cui_polar_grid} studied the sparse characteristic
of single-carrier near-field channels and proposed a polar-domain
sparse representation. However, with multi-carrier and spatial non-stationary,
the sparse representation is quite different.

We first introduce a polar-domain grid of $Q_{1}$ points, where the
sampling points $\left\{ \bar{\vartheta}_{q_{1}},\bar{r}_{q_{1}}\right\} _{q_{1}=1}^{Q_{1}}$
are generated using Algorithm 1 in \cite{Cui_polar_grid}. Then, we
introduce a delay-domain grid of $Q_{2}$ points, such that the sampling
delay points $\left\{ \bar{\tau}_{q_{2}}\right\} _{q_{2}=1}^{Q_{2}}$
are uniformly within $\left[0,\tau_{\textrm{max}}\right]$, where
$\tau_{\textrm{max}}$ is the delay spread. Based on these, the initial
polar-delay domain grid is defined as
\begin{align}
\bar{\mathbf{\Xi}}\triangleq & \left[\left[\bar{\vartheta}_{1},\bar{r}_{1},\bar{\tau}_{1}\right];\ldots;\left[\bar{\vartheta}_{1},\bar{r}_{1},\bar{\tau}_{Q_{2}}\right];\ldots\right.\nonumber \\
 & \left[\bar{\vartheta}_{q_{1}},\bar{r}_{q_{1}},\bar{\tau}_{1}\right];\ldots;\left[\bar{\vartheta}_{q_{1}},\bar{r}_{q_{1}},\bar{\tau}_{Q_{2}}\right];\ldots\nonumber \\
 & \left.\left[\bar{\vartheta}_{Q_{1}},\bar{r}_{Q_{1}},\bar{\tau}_{1}\right];\ldots;\left[\bar{\vartheta}_{Q_{1}},\bar{r}_{Q_{1}},\bar{\tau}_{Q_{2}}\right]\right].
\end{align}
It is obvious that the fixed grid $\bar{\mathbf{\Xi}}$ of $Q=Q_{1}\times Q_{2}$
points usually cannot cover the true angle, distance, and delay parameters
of scatterers. And thus, the channel estimation performance is limited
by the grid interval \cite{YC_beam_squint}. To achieve high-accuracy
channel tracking, we introduce a dynamic polar-delay domain grid in
each frame, denoted by $\mathbf{\Xi}^{\left(t\right)}\triangleq\left[\boldsymbol{\vartheta}^{\left(t\right)},\boldsymbol{r}^{\left(t\right)},\boldsymbol{\tau}^{\left(t\right)}\right],\forall t$,
where $\boldsymbol{\vartheta}^{\left(t\right)}\triangleq\left[\vartheta_{1}^{\left(t\right)},\ldots,\vartheta_{Q}^{\left(t\right)}\right]^{T}$,
$\boldsymbol{r}^{\left(t\right)}\triangleq\left[r_{1}^{\left(t\right)},\ldots,r_{Q}^{\left(t\right)}\right]^{T}$,
and $\boldsymbol{\tau}^{\left(t\right)}\triangleq\left[\tau_{1}^{\left(t\right)},\ldots,\tau_{Q}^{\left(t\right)}\right]^{T}$
represent the angle, distance, and delay grid vectors, respectively,
at time $t$. We can initialize the dynamic grid by $\mathbf{\Xi}^{\left(0\right)}=\bar{\mathbf{\Xi}}$.

Then, the polar-delay domain basis is obtained as
\begin{equation}
\mathbf{B}\left(\mathbf{\Xi}^{\left(t\right)}\right)\triangleq\left[\boldsymbol{b}\left(\vartheta_{1}^{\left(t\right)},r_{1}^{\left(t\right)},\tau_{1}^{\left(t\right)}\right),\ldots,\boldsymbol{b}\left(\vartheta_{Q}^{\left(t\right)},r_{Q}^{\left(t\right)},\tau_{Q}^{\left(t\right)}\right)\right],
\end{equation}
with $\boldsymbol{b}\left(\vartheta_{q}^{\left(t\right)},r_{q}^{\left(t\right)},\tau_{q}^{\left(t\right)}\right)\in\mathbb{C}^{MN\times1},\forall q$
given by
\begin{equation}
\boldsymbol{b}\left(\vartheta_{q}^{\left(t\right)},r_{q}^{\left(t\right)},\tau_{q}^{\left(t\right)}\right)=\boldsymbol{a}\left(\vartheta_{q}^{\left(t\right)},r_{q}^{\left(t\right)}\right)\otimes\boldsymbol{d}\left(\tau_{q}^{\left(t\right)}\right),
\end{equation}
where $\boldsymbol{d}\left(\tau\right)\triangleq\left[e^{-j2\pi f_{0}\tau},\ldots,e^{-j2\pi Nf_{0}\tau}\right]^{T}\in\mathbb{C}^{N\times1}$
represents the delay response vector.

Furthermore, we introduce a VR dictionary corresponding to the polar-delay
domain basis, denoted by $\mathbf{U}^{\left(t\right)}=\left[\boldsymbol{u}_{1}^{\left(t\right)},\ldots,\boldsymbol{u}_{Q}^{\left(t\right)}\right]\in\mathbb{C}^{M\times Q}$,
where $\boldsymbol{u}_{q}^{\left(t\right)}$ represents the VR of
the scatterer lying around the $q\textrm{-th}$ polar-delay domain
grid point. Then, the sparse representation of the channel vector
in (\ref{eq:y_tilde(t)}) can be obtained as
\begin{equation}
\boldsymbol{h}^{\left(t\right)}=\underbrace{\left[\mathbf{B}\left(\mathbf{\Xi}^{\left(t\right)}\right)\odot\left(\mathbf{U}^{\left(t\right)}\otimes\boldsymbol{1}_{N\times1}\right)\right]}_{\mathbf{F}\left(\mathbf{\Xi}^{\left(t\right)},\mathbf{U}^{\left(t\right)}\right)}\boldsymbol{x}^{\left(t\right)},\label{eq:h(t)_sparse}
\end{equation}
where $\boldsymbol{x}^{\left(t\right)}\in\mathbb{C}^{Q\times1}$ is
called the polar-delay domain sparse channel vector and $\mathbf{F}\left(\mathbf{\Xi}^{\left(t\right)},\mathbf{U}^{\left(t\right)}\right)$
is the transform matrix. The $q\textrm{-th}$ element of $\boldsymbol{x}^{\left(t\right)}$,
denoted by $x_{q}^{\left(t\right)}$, represents the complex gain
of the channel path with angle $\vartheta_{q}^{\left(t\right)}$,
distance $r_{q}^{\left(t\right)}$, and delay $\tau_{q}^{\left(t\right)}$.

The sparse representation in (\ref{eq:h(t)_sparse}) is a generalization
of the model in \cite{Cui_polar_grid}. In particular, when $\mathbf{U}^{\left(t\right)}=\boldsymbol{1}_{M\times Q}$,
(\ref{eq:h(t)_sparse}) gives the sparse representation of multi-carrier
spatial stationary channels. Further with $N=1$, (\ref{eq:h(t)_sparse})
reduced to the model in \cite{Cui_polar_grid}.

\subsection{Sparse Prior Model}

The sparse prior model is essential to capture the specific sparse
structure and perform sparse signal recovery. In this subsection,
we first introduce a three-layer Markov prior model to capture the
dynamic sparsity of the polar-delay domain sparse channel vectors.
Then, a hierarchical 2D Markov model is designed to exploit the clustered
sparsity of VRs over antennas and correlation over time.

\subsubsection{Three-layer Markov Prior Model for Channels}

\begin{figure}[t]
\centering{}\includegraphics[width=75mm]{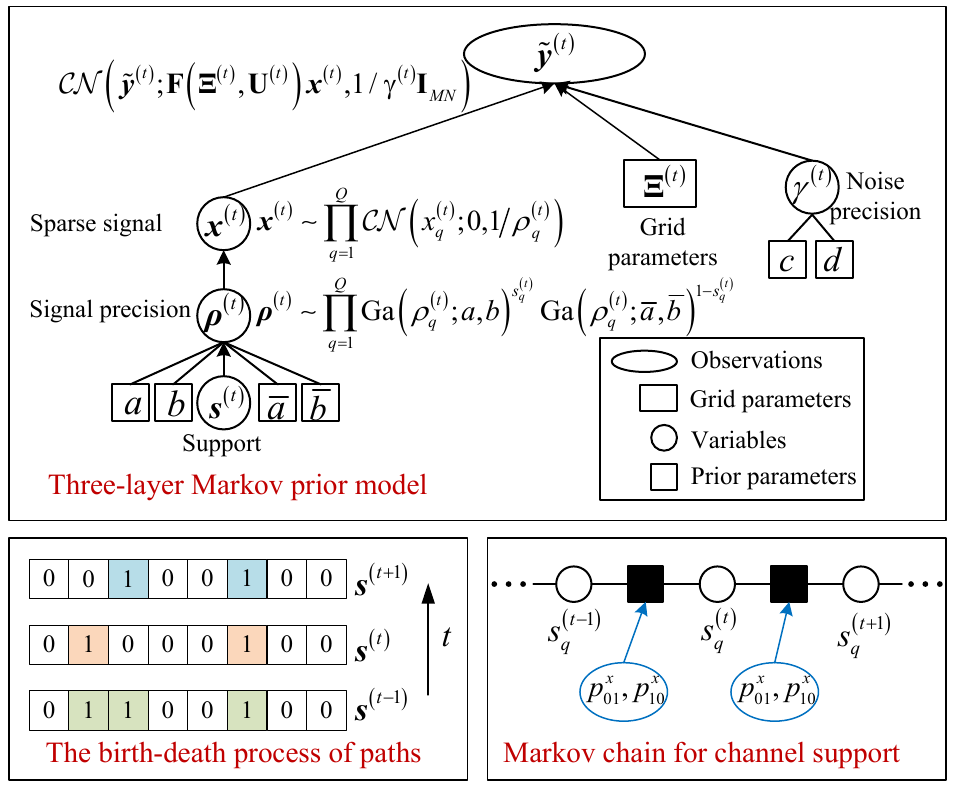}\caption{\label{fig:three_layer_Markov_prior}Three-layer Markov prior model
to exploit the dynamic sparsity of polar-delay domain channels.}
\vspace{-5mm}
\end{figure}

The three-layer Markov prior model is used to describe the sparse
structure of $\boldsymbol{x}^{\left(t\right)},\forall t$, as illustrated
in Fig. \ref{fig:three_layer_Markov_prior}. Define $\boldsymbol{\rho}^{\left(t\right)}\triangleq\left[\rho_{1}^{\left(t\right)},\ldots,\rho_{Q}^{\left(t\right)}\right]^{T}$
and $\boldsymbol{s}^{\left(t\right)}\triangleq\left[s_{1}^{\left(t\right)},\ldots,s_{Q}^{\left(t\right)}\right]^{T}$
as the precision and support vector of $\boldsymbol{x}^{\left(t\right)}$,
respectively, where $1/\rho_{q}^{\left(t\right)}$ is the variance
of $x_{q}^{\left(t\right)}$ and $s_{q}^{\left(t\right)}\in\left\{ 0,1\right\} $
indicates $x_{q}^{\left(t\right)}$ is a non-zero element or not.
Denote the time series $\left\{ \boldsymbol{x}^{\left(1\right)},...,\boldsymbol{x}^{\left(T\right)}\right\} $
as $\boldsymbol{x}^{\left(1:T\right)}$ (same for $\boldsymbol{\rho}^{\left(1:T\right)}$
and $\boldsymbol{s}^{\left(1:T\right)}$), where $T$ is the total
number of frames. The joint distribution of $\boldsymbol{x}^{\left(1:T\right)}$,
$\boldsymbol{\rho}^{\left(1:T\right)}$, and $\boldsymbol{s}^{\left(1:T\right)}$
is expressed as
\begin{align}
 & p\left(\boldsymbol{x}^{\left(1:T\right)},\boldsymbol{\rho}^{\left(1:T\right)},\boldsymbol{s}^{\left(1:T\right)}\right)\nonumber \\
= & \underbrace{p\left(\boldsymbol{s}^{\left(1:T\right)}\right)}_{\textrm{Support}}\prod_{t=1}^{T}\underbrace{p\left(\boldsymbol{\rho}^{\left(t\right)}\mid\boldsymbol{s}^{\left(t\right)}\right)}_{\textrm{Precision}}\prod_{t=1}^{T}\underbrace{p\left(\boldsymbol{x}^{\left(t\right)}\mid\boldsymbol{\rho}^{\left(t\right)}\right)}_{\textrm{Sparse\ signal}}.
\end{align}

Since the scattering environment changes slowly over time, there exists
a strong temporal correlation among support vectors. To be more specific,
if $s_{q}^{\left(t-1\right)}=1$, then $s_{q}^{\left(t\right)}$ is
also equal to 1 with high probability. Therefore, we use a Markov
chain to model support vectors,
\begin{equation}
p\left(\boldsymbol{s}^{\left(1:T\right)}\right)=\prod_{q=1}^{Q}\left(p\left(s_{q}^{\left(1\right)}\right)\prod_{t=2}^{T}p\left(s_{q}^{\left(t\right)}\mid s_{q}^{\left(t-1\right)}\right)\right),\label{eq:p(s(1:T))}
\end{equation}
with the transition probability given by
\begin{subequations}
\begin{align}
p_{01}^{x} & =p\left(s_{q}^{\left(t\right)}=1\mid s_{q}^{\left(t-1\right)}=0\right),\\
p_{10}^{x} & =p\left(s_{q}^{\left(t\right)}=0\mid s_{q}^{\left(t-1\right)}=1\right).
\end{align}
\end{subequations}
The value of the transition probability reflects the strength of temporal
correlation. Specifically, small $p_{01}^{x}$ and $p_{10}^{x}$ imply
a strong temporal correlation, while large $p_{01}^{x}$ and $p_{10}^{x}$
imply the opposite. The steady-state distribution of the Markov chain
is used as the initial distribution, i.e., $p\left(s_{q}^{\left(1\right)}=1\right)=\frac{p_{01}^{x}}{p_{01}^{x}+p_{10}^{x}}$.

The precision vector is modeled as a Bernoulli-Gamma distribution,
which is given by
\begin{equation}
p\left(\boldsymbol{\rho}^{\left(t\right)}\mid\boldsymbol{s}^{\left(t\right)}\right)=\prod_{q=1}^{Q}\textrm{Ga}\left(\rho_{q}^{\left(t\right)};a,b\right)^{s_{q}^{\left(t\right)}}\textrm{Ga}\left(\rho_{q}^{\left(t\right)};\bar{a},\bar{b}\right)^{1-s_{q}^{\left(t\right)}},
\end{equation}
where $a,b$ and $\bar{a},\bar{b}$ are hyper-parameters of the two
Gamma distributions. $s_{q}^{\left(t\right)}=1$ indicates $x_{q}^{\left(t\right)}$
is non-zero, and thus $a,b$ should be chosen to satisfy $\frac{a}{b}=\mathbb{E}\left(\rho_{q}^{\left(t\right)}\mid s_{q}^{\left(t\right)}=1\right)=\mathcal{O}\left(1\right)$.
On the other hand, $s_{q}^{\left(t\right)}=0$ indicates $x_{q}^{\left(t\right)}$
is zero or close to zero, and thus $\bar{a},\bar{b}$ should be chosen
to satisfy $\frac{\bar{a}}{\bar{b}}=\mathbb{E}\left(\rho_{q}^{\left(t\right)}\mid s_{q}^{\left(t\right)}=0\right)\gg1$.
Note that the Gamma distribution is usually used to model the precision
since it is a conjugate of the Gaussian prior, which facilitates closed-form
Bayesian inference \cite{Tipping_SBL,Ji_SBL,Tzikas_VBI,YC_CL}.

The conditional distribution $p\left(\boldsymbol{x}^{\left(t\right)}\mid\boldsymbol{\rho}^{\left(t\right)}\right)$
is given by
\begin{align}
p\left(\boldsymbol{x}^{\left(t\right)}\mid\boldsymbol{\rho}^{\left(t\right)}\right) & =\prod_{q=1}^{Q}p\left(x_{q}^{\left(t\right)}\mid\rho_{q}^{\left(t\right)}\right)\nonumber \\
 & =\prod_{q=1}^{Q}\mathcal{CN}\left(x_{q}^{\left(t\right)};0,1/\rho_{q}^{\left(t\right)}\right).
\end{align}
In addition, a Gamma distribution with hyper-parameters $c$ and $d$
is assumed as the prior for the noise precision,
\begin{equation}
p\left(\gamma^{\left(t\right)}\right)=\textrm{Ga}\left(\gamma^{\left(t\right)};c,d\right).
\end{equation}
The above three-layer Markov prior model is an application of the
existing hierarchical prior model in \cite{LiuAn_CE_Turbo_VBI}, which
has been verified to be robust w.r.t. the imperfect prior information
in practice. Besides, the Markov chain in (\ref{eq:p(s(1:T))}) enables
the prior model to describe the birth-death process of channel paths
over time.

\subsubsection{Hierarchical 2D Markov Model for VRs}

\begin{figure}[t]
\begin{centering}
\includegraphics[width=1\columnwidth]{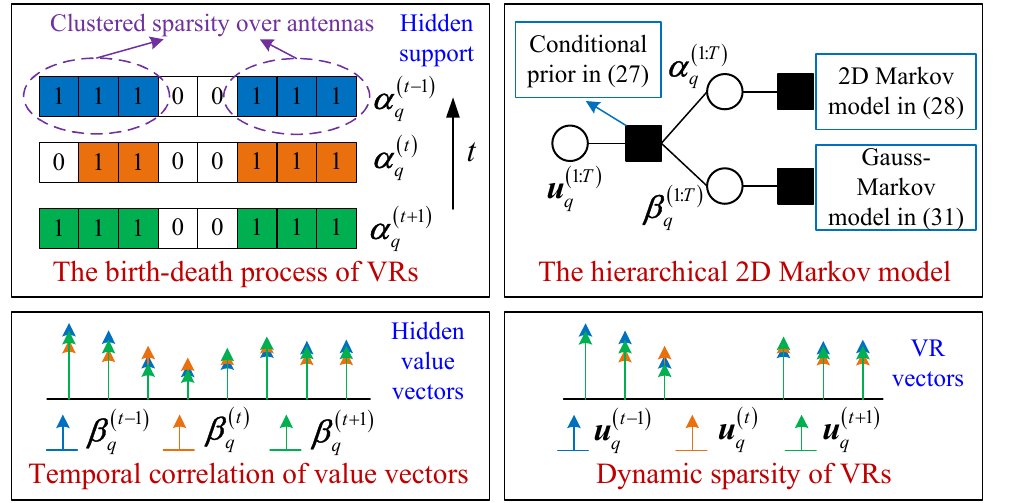}\caption{\label{fig:2D_MM_VR}Hierarchical 2D Markov model to exploit the dynamic
sparsity of VRs.}
\par\end{centering}
\centering{}\vspace{-5mm}
\end{figure}

The VR of scatterers has the following three characteristics: 1) the
visible antennas of each scatterer are concentrated on a few clusters
\cite{Vincent_sparse_modeling,Tang_sparse_modeling}, and thus the
VR vector exhibits a clustered sparsity; 2) similar to the channel
paths, the birth-death process of VRs also exists; 3) the received
power of visible antennas usually change smoothly over time. To exploit
these, we propose a hierarchical 2D Markov model with two hidden random
processes, as illustrated in Fig. \ref{fig:2D_MM_VR}.

Specifically, we define a binary vector $\boldsymbol{\alpha}_{q}^{\left(t\right)}\triangleq\left[\alpha_{q,1}^{\left(t\right)},\ldots,\alpha_{q,M}^{\left(t\right)}\right]^{T}$
as the hidden support vector of $\boldsymbol{u}_{q}^{\left(t\right)}$,
where $\alpha_{q,m}^{\left(t\right)}=1$ indicates the $m\textrm{-th}$
antenna is visible to the scatterer lying in the $q\textrm{-th}$
polar-delay domain grid, while $\alpha_{q,m}^{\left(t\right)}=0$
indicates the opposite. Let $\boldsymbol{\beta}_{q}^{\left(t\right)}\triangleq\left[\beta_{q,1}^{\left(t\right)},\ldots,\beta_{q,M}^{\left(t\right)}\right]^{T}$
represent the hidden value vector of $\boldsymbol{u}_{q}^{\left(t\right)}$,
where $\beta_{q,m}^{\left(t\right)}=u_{q,m}^{\left(t\right)}$ if
$\alpha_{q,m}^{\left(t\right)}=1$. Based on these, the dynamic VR
is modeled as
\begin{equation}
u_{q,m}^{\left(t\right)}=\alpha_{q,m}^{\left(t\right)}\beta_{q,m}^{\left(t\right)},\forall q,\forall m,\forall t.\label{eq:u=002192alpha,beta}
\end{equation}
Then, the joint distribution of $\boldsymbol{u}_{q}^{\left(1:T\right)}$,
$\boldsymbol{\alpha}_{q}^{\left(1:T\right)}$, and $\boldsymbol{\beta}_{q}^{\left(1:T\right)}$
can be expressed as
\begin{align}
 & p\left(\boldsymbol{u}_{q}^{\left(1:T\right)},\boldsymbol{\alpha}_{q}^{\left(1:T\right)},\boldsymbol{\beta}_{q}^{\left(1:T\right)}\right)\nonumber \\
= & \underbrace{p\left(\boldsymbol{\alpha}_{q}^{\left(1:T\right)}\right)}_{\textrm{Hidden\ support}}\underbrace{p\left(\boldsymbol{\beta}_{q}^{\left(1:T\right)}\right)}_{\textrm{Hidden\ value}}\prod_{t=1}^{T}\underbrace{p\left(\boldsymbol{u}_{q}^{\left(t\right)}\mid\boldsymbol{\alpha}_{q}^{\left(t\right)},\boldsymbol{\beta}_{q}^{\left(t\right)}\right)}_{\textrm{VR\ vector}}.\label{eq:p(u,alpha,beta)}
\end{align}
According to (\ref{eq:u=002192alpha,beta}), the conditional distribution
is given by
\begin{align}
p\left(\boldsymbol{u}_{q}^{\left(t\right)}\mid\boldsymbol{\alpha}_{q}^{\left(t\right)},\boldsymbol{\beta}_{q}^{\left(t\right)}\right) & =\prod_{m=1}^{M}p\left(u_{q,m}^{\left(t\right)}\mid\alpha_{q,m}^{\left(t\right)},\beta_{q,m}^{\left(t\right)}\right)\nonumber \\
 & =\prod_{m=1}^{M}\delta\left(u_{q,m}^{\left(t\right)}-\alpha_{q,m}^{\left(t\right)}\beta_{q,m}^{\left(t\right)}\right),\label{eq:p(u_q1(t)|alpha_q1(t),beta_q1(t))}
\end{align}
where $\delta\left(\cdot\right)$ is the Dirac Delta function.

The hidden support vectors exhibit a 2D clustered sparsity over antennas
and time. Specifically, if $\alpha_{q,m}^{\left(t-1\right)}=1$ or
$\alpha_{q,m-1}^{\left(t\right)}=1$, there is a higher probability
that $\alpha_{q,m}^{\left(t\right)}=1$. Therefore, we use a 2D Markov
prior to model the hidden support vectors \cite{Fornasini_2D_MM,Xu_alternating_MAP},
\begin{align}
p\left(\boldsymbol{\alpha}_{q}^{\left(1:T\right)}\right)= & p\left(\alpha_{q,1}^{\left(1\right)}\right)\prod_{t=2}^{T}\prod_{m=1}^{M}p\left(\alpha_{q,m}^{\left(t\right)}\mid\alpha_{q,m}^{\left(t-1\right)}\right)\nonumber \\
 & \times\prod_{t=1}^{T}\prod_{m=2}^{M}p\left(\alpha_{q,m}^{\left(t\right)}\mid\alpha_{q,m-1}^{\left(t\right)}\right),\forall q,
\end{align}
with the transition probability given by
\begin{subequations}
\begin{align}
p_{01}^{T} & =p\left(\alpha_{q,m}^{\left(t\right)}=1\mid\alpha_{q,m}^{\left(t-1\right)}=0\right),\label{eq:p01_T}\\
p_{10}^{T} & =p\left(\alpha_{q,m}^{\left(t\right)}=0\mid\alpha_{q,m}^{\left(t-1\right)}=1\right),\label{eq:p10_T}\\
p_{01}^{S} & =p\left(\alpha_{q,m}^{\left(t\right)}=1\mid\alpha_{q,m-1}^{\left(t\right)}=0\right),\label{eq:p01_S}\\
p_{10}^{S} & =p\left(\alpha_{q,m}^{\left(t\right)}=0\mid\alpha_{q,m-1}^{\left(t\right)}=1\right).\label{eq:p10_S}
\end{align}
\end{subequations}
The initial distribution $p\left(\alpha_{q,1}^{\left(1\right)}=1\right)=\kappa$
is set to be the steady-state distribution of the 2D Markov model
\cite{Fornasini_2D_MM}.

The specific sparse structure of hidden support vectors is affected
by the value of $\left\{ p_{01}^{T},p_{10}^{T},p_{01}^{S},p_{10}^{S}\right\} $.
Specifically, a smaller $p_{01}^{S}$ leads to a larger average cluster
size of visible antennas, a smaller $p_{01}^{S}$ implies a larger
average gap between two adjacent visible clusters, and smaller $p_{01}^{T}$
and $p_{10}^{T}$ indicate a stronger temporal correlation of VRs.
Therefore, the 2-D Markov model is very suitable to describe the dynamic
sparsity of VRs.

As shown in Fig. \ref{fig:2D_MM_VR}, the received power of antennas
change smoothly over time. And thus, we use a spatially independent
steady-state Gauss-Markov process to model the temporal correlation
of the hidden value vectors \cite{Ziniel_AMP_Tracking},
\begin{equation}
\boldsymbol{\beta}_{q,m}^{\left(t\right)}=\left(1-\varepsilon\right)\left(\boldsymbol{\beta}_{q,m}^{\left(t-1\right)}-\zeta\right)+\varepsilon\varpi_{q,m}^{\left(t\right)}+\zeta,
\end{equation}
where $\varepsilon\in\left[0,1\right]$ controls the strength of temporal
correlation, $\varpi_{q,m}^{\left(t\right)}\sim\mathcal{N}\left(0,\sigma^{2}\right)$
is the Gaussian perturbation with variance $\sigma^{2}$, and $\zeta\in\mathbb{R}+$
is the mean of the random process. Then, $p\left(\boldsymbol{\beta}_{q}^{\left(1:T\right)}\right),\forall q$
is given by
\begin{equation}
p\left(\boldsymbol{\beta}_{q}^{\left(1:T\right)}\right)=\prod_{m=1}^{M}\left(p\left(\boldsymbol{\beta}_{q,m}^{\left(1\right)}\right)\prod_{t=2}^{T}p\left(\boldsymbol{\beta}_{q,m}^{\left(t\right)}\mid\boldsymbol{\beta}_{q,m}^{\left(t-1\right)}\right)\right),\label{eq:p(beta_q^(1:T))}
\end{equation}
where the initial distribution is $\boldsymbol{\beta}_{q,m}^{\left(1\right)}\sim\mathcal{N}\left(\boldsymbol{\beta}_{q,m}^{\left(1\right)};\zeta,\frac{\varepsilon\sigma^{2}}{2-\varepsilon}\right)$,
and the conditional distribution is given by
\[
p\left(\boldsymbol{\beta}_{q,m}^{\left(t\right)}\mid\boldsymbol{\beta}_{q,m}^{\left(t-1\right)}\right)=\mathcal{N}\left(\boldsymbol{\beta}_{q,m}^{\left(t\right)};\left(1-\varepsilon\right)\boldsymbol{\beta}_{q,m}^{\left(t-1\right)}+\varepsilon\zeta,\varepsilon^{2}\sigma^{2}\right).
\]
Although the prior distribution in (\ref{eq:p(beta_q^(1:T))}) cannot
ensure $u_{q,m}^{\left(t\right)}$ is non-negative, the MAP estimate
of $u_{q,m}^{\left(t\right)}$ must satisfy $\hat{u}_{q,m}^{\left(t\right)}\geq0$,
since it is obtained based on both the estimated posterior distribution
and the constraint $u_{q,m}^{\left(t\right)}\geq0$, as presented
in (\ref{eq:Uq1_hat_(t)}).

\subsection{Channel Tracking Problem Formulation}

Based on the polar-delay domain sparse representation in (\ref{eq:h(t)_sparse}),
the received signal in (\ref{eq:y_tilde(t)}) can be reformulated
into the following bilinear measurement process:
\begin{equation}
\tilde{\boldsymbol{y}}^{\left(t\right)}=\mathbf{F}\left(\mathbf{\Xi}^{\left(t\right)},\mathbf{U}^{\left(t\right)}\right)\boldsymbol{x}^{\left(t\right)}+\tilde{\boldsymbol{z}}^{\left(t\right)},t=1,\ldots,T,\label{eq:y_tilde(t)_final}
\end{equation}
where the observation $\tilde{\boldsymbol{y}}^{\left(t\right)}$ is
linear w.r.t. $\boldsymbol{x}^{\left(t\right)}$ and $\mathbf{U}^{\left(t\right)}$.
At each time $t$, the goal is to obtain the MAP estimate of the channel
vector $\boldsymbol{x}^{\left(t\right)}$, the VR $\mathbf{U}^{\left(t\right)}$,
and the grid parameters $\mathbf{\Xi}^{\left(t\right)}$ based on
all the previous observations up to time $t$, i.e.,
\begin{equation}
\boldsymbol{x}_{\textrm{MAP}}^{\left(t\right)},\mathbf{U}_{\textrm{MAP}}^{\left(t\right)},\mathbf{\Xi}_{\textrm{MAP}}^{\left(t\right)}=\underset{\mathbf{\Theta}^{\left(t\right)},\mathbf{\Xi}^{\left(t\right)}}{\arg\max}\ln p\left(\tilde{\boldsymbol{y}}^{\left(1:t\right)},\mathbf{\Theta}^{\left(t\right)};\mathbf{\Xi}^{\left(t\right)}\right),\label{eq:MAP_problem}
\end{equation}
where $\mathbf{\Theta}^{\left(t\right)}\triangleq\left\{ \boldsymbol{x}^{\left(t\right)},\boldsymbol{\rho}^{\left(t\right)},\boldsymbol{s}^{\left(t\right)},\gamma^{\left(t\right)},\mathbf{U}^{\left(t\right)},\boldsymbol{\alpha}_{1:Q}^{\left(t\right)},\boldsymbol{\beta}_{1:Q}^{\left(t\right)}\right\} $
is the collection of random variables at time $t$.

One possible solution is to store all the available observations $\tilde{\boldsymbol{y}}^{\left(1:t\right)}$
and perform a joint estimation of $\boldsymbol{x}^{\left(t\right)}$,
$\mathbf{U}^{\left(t\right)}$, and $\mathbf{\Xi}^{\left(t\right)}$
at each time $t$. However, the memory cost and computational complexity
of such a brute-force solution would become unacceptable for a large
$t$. To address this challenge, we decompose and approximate the
joint distribution $p\left(\tilde{\boldsymbol{y}}^{\left(1:t\right)},\mathbf{\Theta}^{\left(t\right)};\mathbf{\Xi}^{\left(t\right)}\right)$
to make it only involve the probability density function (PDF) of
the current random variables $\mathbf{\Theta}^{\left(t\right)}$,
the current observation $\tilde{\boldsymbol{y}}^{\left(t\right)}$,
and the messages passed from time $t-1$, based on which a more efficient
algorithm can be designed. In particular, we have
\begin{align}
 & p\left(\tilde{\boldsymbol{y}}^{\left(1:t\right)},\mathbf{\Theta}^{\left(t\right)};\mathbf{\Xi}^{\left(t\right)}\right)\nonumber \\
\propto & \check{p}^{\left(t\right)}\sum_{\boldsymbol{s}^{(t-1)}}\sum_{\boldsymbol{\alpha}_{1:Q}^{\left(t-1\right)}}\int_{\boldsymbol{\beta}_{1:Q}^{\left(t-1\right)}}p\left(\boldsymbol{s}^{\left(t-1\right)},\boldsymbol{\alpha}_{1:Q}^{\left(t-1\right)},\boldsymbol{\beta}_{1:Q}^{\left(t-1\right)}\mid\tilde{\boldsymbol{y}}^{\left(1:t-1\right)}\right)\nonumber \\
 & \times p\left(\boldsymbol{s}^{\left(t\right)}\mid\boldsymbol{s}^{\left(t-1\right)}\right)\prod_{q=1}^{Q}p\left(\boldsymbol{\alpha}_{q}^{\left(t\right)}|\boldsymbol{\alpha}_{q}^{\left(t-1\right)}\right)p\left(\boldsymbol{\beta}_{q}^{\left(t\right)}|\boldsymbol{\beta}_{q}^{\left(t-1\right)}\right)\nonumber \\
\approx & \check{p}^{\left(t\right)}\underbrace{\sum_{\boldsymbol{s}^{(t-1)}}q\left(\boldsymbol{s}^{\left(t-1\right)}\right)p\left(\boldsymbol{s}^{\left(t\right)}|\boldsymbol{s}^{\left(t-1\right)}\right)}_{\hat{p}\left(\boldsymbol{s}^{\left(t\right)}\right)}\nonumber \\
 & \times\prod_{q=1}^{Q}\underbrace{\sum_{\boldsymbol{\alpha}_{q}^{\left(t-1\right)}}q\left(\boldsymbol{\alpha}_{q}^{\left(t-1\right)}\right)p\left(\boldsymbol{\alpha}_{q}^{\left(t\right)}|\boldsymbol{\alpha}_{q}^{\left(t-1\right)}\right)}_{\hat{p}\left(\boldsymbol{\alpha}_{q}^{\left(t\right)}\right)}\nonumber \\
 & \times\prod_{q=1}^{Q}\underbrace{\int_{\boldsymbol{\beta}_{q}^{\left(t-1\right)}}q\left(\boldsymbol{\beta}_{q}^{\left(t-1\right)}\right)p\left(\boldsymbol{\beta}_{q}^{\left(t\right)}|\boldsymbol{\beta}_{q}^{\left(t-1\right)}\right)}_{\hat{p}\left(\boldsymbol{\beta}_{q}^{\left(t\right)}\right)}\nonumber \\
\triangleq & \check{p}\left(\tilde{\boldsymbol{y}}^{\left(t\right)},\mathbf{\Theta}^{\left(t\right)};\mathbf{\Xi}^{\left(t\right)}\right),\label{eq:p(y^(1:t),alphe^(t),beta^(t))}
\end{align}
where $\check{p}^{\left(t\right)}$ is given by
\begin{align}
 & \check{p}^{\left(t\right)}\triangleq p\left(\tilde{\boldsymbol{y}}^{\left(t\right)}\mid\boldsymbol{x}^{\left(t\right)},\mathbf{U}^{\left(t\right)},\gamma^{\left(t\right)};\mathbf{\Xi}^{\left(t\right)}\right)p\left(\gamma^{\left(t\right)}\right)\nonumber \\
 & \times p\left(\boldsymbol{x}^{\left(t\right)}|\boldsymbol{\rho}^{\left(t\right)}\right)p\left(\boldsymbol{\rho}^{\left(t\right)}|\boldsymbol{s}^{\left(t\right)}\right)\prod_{q=1}^{Q}p\left(\boldsymbol{u}_{q}^{\left(t\right)}\mid\boldsymbol{\alpha}_{q}^{\left(t\right)},\boldsymbol{\beta}_{q}^{\left(t\right)}\right),\label{eq:p(y_tilde^(t),theta^(t))}
\end{align}
with the likelihood function in (\ref{eq:p(y_tilde^(t),theta^(t))})
given by $\mathcal{CN}\left(\tilde{\boldsymbol{y}}^{\left(t\right)};\mathbf{F}\left(\mathbf{\Xi}^{\left(t\right)},\mathbf{U}^{\left(t\right)}\right)\boldsymbol{x}^{\left(t\right)},1/\gamma^{\left(t\right)}\mathbf{I}_{MN}\right)$.
Note that we cannot obtain the exact posterior distribution $p\left(\boldsymbol{s}^{\left(t-1\right)},\boldsymbol{\alpha}_{1:Q}^{\left(t-1\right)},\boldsymbol{\beta}_{1:Q}^{\left(t-1\right)}\mid\tilde{\boldsymbol{y}}^{\left(1:t-1\right)}\right)$
in (\ref{eq:p(y^(1:t),alphe^(t),beta^(t))}) since the factor graph
corresponding to the joint distribution $p\left(\tilde{\boldsymbol{y}}^{\left(1:t-1\right)},\mathbf{\Theta}^{\left(1:t-1\right)};\mathbf{\Xi}^{\left(1:t-1\right)}\right)$
contains loops. Therefore, we turn to calculate the approximate posterior
of $\boldsymbol{s}^{\left(t-1\right)}$, $\boldsymbol{\alpha}_{q}^{\left(t-1\right)}$,
and $\boldsymbol{\beta}_{q}^{\left(t-1\right)}$, denoted by $q\left(\boldsymbol{s}^{\left(t-1\right)}\right)$,
$q\left(\boldsymbol{\alpha}_{q}^{\left(t-1\right)}\right)$, and $q\left(\boldsymbol{\beta}_{q}^{\left(t-1\right)}\right)$,
respectively. Then $\hat{p}\left(\boldsymbol{s}^{\left(t\right)}\right)$,
$\hat{p}\left(\boldsymbol{\alpha}_{q}^{\left(t\right)}\right)$, and
$\hat{p}\left(\boldsymbol{\beta}_{q}^{\left(t\right)}\right)$ can
be viewed as the effective prior information for $\boldsymbol{s}^{\left(t\right)}$,
$\boldsymbol{\alpha}_{q}^{\left(t\right)}$, and $\boldsymbol{\beta}_{q}^{\left(t\right)}$
obtained by prediction from time $t-1$, which summarize all information
contributed by $\tilde{\boldsymbol{y}}^{\left(1:t-1\right)}$. Using
(\ref{eq:p(y^(1:t),alphe^(t),beta^(t))}), the MAP problem in (\ref{eq:MAP_problem})
can be simplified into
\begin{align}
\boldsymbol{x}_{\textrm{MAP}}^{\left(t\right)},\mathbf{U}_{\textrm{MAP}}^{\left(t\right)},\mathbf{\Xi}_{\textrm{MAP}}^{\left(t\right)}= & \underset{\mathbf{\Theta}^{\left(t\right)},\mathbf{\Xi}^{\left(t\right)}}{\arg\max}\ln\check{p}\left(\tilde{\boldsymbol{y}}^{\left(t\right)},\mathbf{\Theta}^{\left(t\right)};\mathbf{\Xi}^{\left(t\right)}\right).\label{eq:MAP_problem_simplified}
\end{align}
However, it is still intractable to directly solve the MAP estimation
problem in (\ref{eq:MAP_problem_simplified}) since different variables
and their latent variables have quite different priors and they are
complicatedly coupled with the likelihood function. In the next section,
we shall propose the DA-MAP framework to overcome this challenge.

\section{Dynamic Alternating MAP Framework}

\subsection{Outline of DA-MAP Framework}

\begin{figure}[t]
\centering{}\includegraphics[width=0.85\columnwidth]{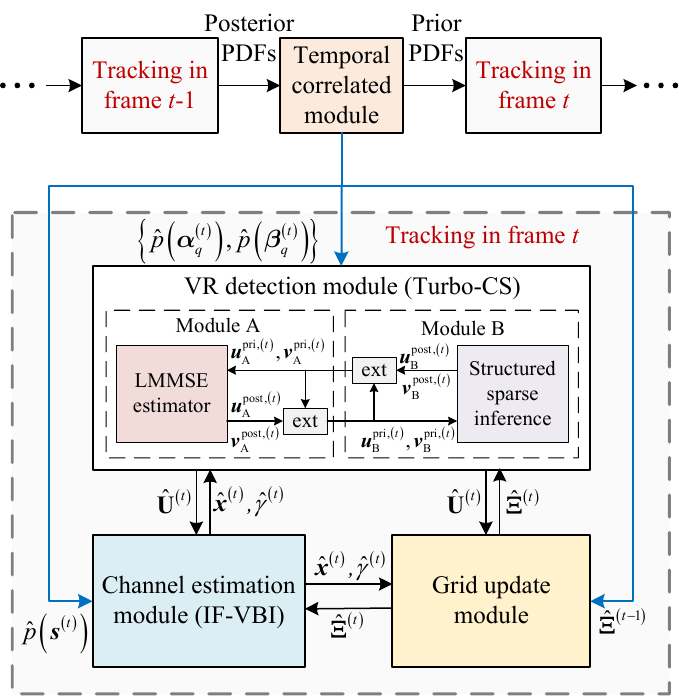}\caption{\label{fig:DA-MAP}The DA-MAP algorithmic framework with its four
basic modules.}
\vspace{-2mm}
\end{figure}

The dynamic alternating MAP framework adopts the idea of alternating
optimization to solve the MAP problem in (\ref{eq:MAP_problem_simplified}).
As illustrated in Fig. \ref{fig:DA-MAP}, the DA-MAP consists of four
basic modules: channel estimation module, VR detection module, grid
update module, and temporal correlated module. In the following, we
give a brief introduction of each basic module.
\begin{itemize}
\item \textbf{Channel estimation module:} The low-complexity IF-VBI estimator
is used to achieve channel estimation. Specifically, given $\hat{p}\left(\boldsymbol{s}^{\left(t\right)}\right)$
from time $t-1$, $\hat{\mathbf{U}}^{\left(t\right)}$ from VR detection
module, and $\hat{\mathbf{\Xi}}^{\left(t\right)}$ from grid update
module, the IF-VBI performs Bayesian inference to compute the posterior
distribution of $\boldsymbol{x}^{\left(t\right)}$, $\boldsymbol{\rho}^{\left(t\right)}$,
$\boldsymbol{s}^{\left(t\right)}$, and $\gamma^{\left(t\right)}$
in an alternative way. 
\item \textbf{VR detection module:} It is the Turbo-CS algorithm that combines
the linear minimum-mean-square-error (LMMSE) module and the structured
sparse inference module via the turbo approach. Given $\hat{p}\left(\boldsymbol{\alpha}_{q}^{\left(t\right)}\right)$
and $\hat{p}\left(\boldsymbol{\beta}_{q}^{\left(t\right)}\right)$
from time $t-1$, $\hat{\boldsymbol{x}}^{\left(t\right)}$and $\hat{\gamma}^{\left(t\right)}$
from channel estimation module, and $\hat{\mathbf{\Xi}}^{\left(t\right)}$
from grid update module, the VR detection module calculates the marginal
posterior distribution of $\boldsymbol{u}_{q}^{\left(t\right)}$,
$\boldsymbol{\alpha}_{q}^{\left(t\right)}$, and $\boldsymbol{\beta}_{q}^{\left(t\right)}$,
$\forall q$.
\item \textbf{Grid update module:} It uses the gradient ascent method to
refine the polar-delay domain grid. Given $\hat{\mathbf{\Xi}}^{\left(t-1\right)}$
from time $t-1$, $\hat{\boldsymbol{x}}^{\left(t\right)}$and $\hat{\gamma}^{\left(t\right)}$
from channel estimation module, and $\hat{\mathbf{U}}^{\left(t\right)}$
from VR detection module, calculate the gradient of $\ln p\left(\tilde{\boldsymbol{y}}^{\left(t\right)}\mid\hat{\boldsymbol{x}}^{\left(t\right)},\hat{\mathbf{U}}^{\left(t\right)},\hat{\gamma}^{\left(t\right)};\mathbf{\Xi}^{\left(t\right)}\right)$,
and then update grid parameters via gradient ascent.
\item \textbf{Temporal correlated module:} Given $q\left(\boldsymbol{s}^{\left(t\right)}\right)$
from channel estimation module, $q\left(\boldsymbol{\alpha}_{q}^{\left(t\right)}\right)$
and $q\left(\boldsymbol{\beta}_{q}^{\left(t\right)}\right)$ from
VR detection module, it calculates $\hat{p}\left(\boldsymbol{s}^{\left(t+1\right)}\right)$,
$\hat{p}\left(\boldsymbol{\alpha}_{q}^{\left(t+1\right)}\right)$,
and $\hat{p}\left(\boldsymbol{\beta}_{q}^{\left(t+1\right)}\right)$
as the prior for time $t+1$.
\end{itemize}

At each time $t$, the channel estimation module, VR detection module,
and grid update module work alternatively until convergence to a stationary
point of (\ref{eq:MAP_problem_simplified}). Then, the temporal correlated
module calculates the prior information for time $t+1$. The proposed
DA-MAP can be viewed as an extension of the alternating MAP framework
in \cite{Xu_alternating_MAP} by: 1) extending from 0/1 VR detection
to continuous VR detection with power change; 2) extending from the
single-frame channel estimation to multi-frame channel tracking. 

\subsection{Channel Estimation Module (IF-VBI)}

Given $\hat{\mathbf{U}}^{\left(t\right)}$ and $\hat{\mathbf{\Xi}}^{\left(t\right)}$,
the transform matrix $\mathbf{F}\left(\hat{\mathbf{\Xi}}^{\left(t\right)},\hat{\mathbf{U}}^{\left(t\right)}\right)$
in (\ref{eq:y_tilde(t)_final}) is fixed. Now the observation in (\ref{eq:y_tilde(t)_final})
can be simplified into $\tilde{\boldsymbol{y}}^{\left(t\right)}=\mathbf{F}\boldsymbol{x}^{\left(t\right)}+\tilde{\boldsymbol{z}}^{\left(t\right)}$,
where $\hat{\mathbf{\Xi}}^{\left(t\right)}$ and $\hat{\mathbf{U}}^{\left(t\right)}$
are omitted in $\mathbf{F}\left(\hat{\mathbf{\Xi}}^{\left(t\right)},\hat{\mathbf{U}}^{\left(t\right)}\right)$
for simplicity. The sub-problem of channel estimation is a standard
compressive sensing problem that can be solved by many existing CS-based
methods. We adopt the IF-VBI algorithm to achieve a good trade-off
between the estimation performance and computational complexity. 

The IF-VBI avoids the complicated matrix inverse each iteration via
minimizing a relaxed Kullback-Leibler divergence. The prior distribution
$\hat{p}\left(\boldsymbol{s}^{\left(t\right)}\right)$ from time $t-1$
can speed up convergence of the IF-VBI algorithm. Due the the space
limitation, please refer to our previous work in \cite{Xu_alternating_MAP,Xu_Turbo-IFVBI}
for more details of the IF-VBI. 

\subsection{VR Detection Module (Turbo-CS)\label{subsec:VR-Detection-Module}}

Given $\hat{\boldsymbol{x}}^{\left(t\right)}$, $\hat{\gamma}^{\left(t\right)}$,
and $\hat{\mathbf{\Xi}}^{\left(t\right)}$, the observation in (\ref{eq:y_tilde(t)_final})
can be rewritten into
\begin{equation}
\tilde{\boldsymbol{y}}^{\left(t\right)}=\left[\mathbf{B}\odot\left(\mathbf{U}^{\left(t\right)}\otimes\boldsymbol{1}_{N\times1}\right)\right]\hat{\boldsymbol{x}}^{\left(t\right)}+\tilde{\boldsymbol{z}}^{\left(t\right)},
\end{equation}
where $\hat{\mathbf{\Xi}}^{\left(t\right)}$ is omitted in $\mathbf{B}\left(\hat{\mathbf{\Xi}}^{\left(t\right)}\right)$
to simply the notation. Define $\boldsymbol{\Phi}_{m}\triangleq\left\{ \left(m-1\right)N+n\mid n=1,\ldots,N\right\} $,
the received signal of the $m\textrm{-th}$ antenna is given by
\begin{align}
\tilde{\boldsymbol{y}}_{m}^{\left(t\right)} & =\underbrace{\left[\mathbf{B}_{m}\odot\left(\boldsymbol{1}_{N\times1}\left(\hat{\boldsymbol{x}}^{\left(t\right)}\right)^{T}\right)\right]}_{\mathbf{G}_{m}^{\left(t\right)}}\left(\mathbf{U}_{m,:}^{\left(t\right)}\right)^{T}+\tilde{\boldsymbol{z}}_{m}^{\left(t\right)},\label{eq:ym_tilde(t)}
\end{align}
where $\tilde{\boldsymbol{y}}_{m}^{\left(t\right)}\in\mathbb{C}^{N\times1}$
and $\tilde{\boldsymbol{z}}_{m}^{\left(t\right)}\in\mathbb{C}^{N\times1}$
denote the $m\textrm{-th}$ column of $\tilde{\mathbf{Y}}^{\left(t\right)}$
and $\tilde{\mathbf{Z}}^{\left(t\right)}$ in (\ref{eq:Y_tilde(t)}),
respectively, $\mathbf{B}_{m}\triangleq\left[\mathbf{B}_{k,:}\right]_{k\in\boldsymbol{\Phi}_{m}}$
with $\mathbf{B}_{k,:}$ denote the $k\textrm{-th}$ row of $\mathbf{B}$,
and $\mathbf{U}_{m,:}^{\left(t\right)}\triangleq\left[u_{1,m}^{\left(t\right)},\ldots,u_{Q,m}^{\left(t\right)}\right]$
represents the $m\textrm{-th}$ row of $\mathbf{U}^{\left(t\right)}$. 

Note that $\hat{\boldsymbol{x}}^{\left(t\right)}$ has a few non-zero
elements since the number of channel paths is much smaller than the
number of grid points, i.e., $L^{\left(t\right)}\ll Q$. Therefore,
the sensing matrix $\mathbf{G}_{m}^{\left(t\right)}$ also has many
close-to-zero columns and is ill-conditioned, which makes it difficult
to estimate $\mathbf{U}_{m,:}^{\left(t\right)}$ accurately. To address
this issue, we introduce a polar-delay domain filtering method. Specifically,
we compare the energy of each element of $\hat{\boldsymbol{x}}^{\left(t\right)}$
with an energy threshold $\eta>0$. Define $\mathbf{\Omega}^{\left(t\right)}\triangleq\left\{ q\mid\forall\left\Vert \hat{x}_{q}^{\left(t\right)}\right\Vert ^{2}>\eta\right\} $
as the index set of the elements with the energy larger than the threshold.
Then, we only retain the columns indexed by $\mathbf{\Omega}^{\left(t\right)}$
in $\mathbf{G}_{m}^{\left(t\right)}$ and delete other columns that
are close to zero. The sensing matrix after clipping, denoted by $\tilde{\mathbf{G}}_{m}^{\left(t\right)}\in\mathbb{C}^{N\times\left|\mathbf{\Omega}^{\left(t\right)}\right|}$,
is well-conditioned now. Moreover, we only need to estimate $\ddot{\boldsymbol{u}}_{m}^{\left(t\right)}\triangleq\left[u_{q,m}^{\left(t\right)}\right]_{q\in\mathbf{\Omega}^{\left(t\right)}}$
instead of $\mathbf{U}_{m,:}^{\left(t\right)}$. This is because the
energy of $\hat{x}_{q}^{\left(t\right)}$ is close to zero for $q\notin\mathbf{\Omega}^{\left(t\right)}$,
which implies that there is no scatterer lying around the $q\textrm{-th}$
polar-delay domain grid. In this case, we have $u_{q,m}^{\left(t\right)}=0,\forall m$.

Based on the polar-delay domain filtering, the signal model in (\ref{eq:ym_tilde(t)})
is rewritten as
\begin{equation}
\tilde{\boldsymbol{y}}_{m}^{\left(t\right)}=\tilde{\mathbf{G}}_{m}^{\left(t\right)}\ddot{\boldsymbol{u}}_{m}^{\left(t\right)}+\tilde{\boldsymbol{z}}_{m}^{\left(t\right)},\forall m,\label{eq:ym_tilde(t)_comp}
\end{equation}
which is a complex-valued observation model while $\ddot{\boldsymbol{u}}_{m}$
is real-valued. And thus, we transform (\ref{eq:ym_tilde(t)_comp})
into a real-valued one,
\begin{equation}
\boldsymbol{\ddot{y}}_{m}^{\left(t\right)}=\mathbf{\ddot{G}}_{m}^{\left(t\right)}\boldsymbol{\ddot{u}}_{m}^{\left(t\right)}+\boldsymbol{\ddot{z}}_{m}^{\left(t\right)},\forall m,\label{eq:ym_dot_(t)}
\end{equation}
where $\boldsymbol{\ddot{y}}_{m}^{\left(t\right)}\triangleq\left[\Re\left(\boldsymbol{\ddot{y}}_{m}^{\left(t\right)}\right)^{T},\Im\left(\boldsymbol{\ddot{y}}_{m}^{\left(t\right)}\right)^{T}\right]^{T},$
and $\mathbf{\ddot{G}}_{m}^{\left(t\right)}$ and $\boldsymbol{\ddot{z}}_{m}^{\left(t\right)}$
are defined similarly. The sub-problem of VR detection is to obtain
the posterior distribution of $\ddot{\boldsymbol{u}}_{m},\forall m$
based on the linear observations in (\ref{eq:ym_dot_(t)}) and structured
sparse priors. Inspired by the turbo framework \cite{LiuAn_CE_Turbo_CS},
we develop a Turbo-CS algorithm to achieve VR detection in a parallel
fashion.

As shown in Fig. \ref{fig:DA-MAP}, the Turbo-CS has two basic modules:
Module A and Module B. Specifically, Module A performs the LMMSE estimation
based on the observation of each antenna and messages from Module
B, while Module B performs structured sparse inference using the hierarchical
2D Markov prior information and messages from Module A. The two modules
exchange messages with each other until convergence.

\subsubsection{LMMSE in Module A}

The prior distribution of $\boldsymbol{\ddot{u}}_{m}^{\left(t\right)}$
in Module A is $\mathcal{N}\left(\boldsymbol{\ddot{u}}_{m}^{\left(t\right)};\boldsymbol{u}_{\textrm{A},m}^{\textrm{pri},\left(t\right)},\boldsymbol{v}_{\textrm{A},m}^{\textrm{pri,\ensuremath{\left(t\right)}}}\right)$,
where $\boldsymbol{u}_{\textrm{A},m}^{\textrm{pri},\left(t\right)}$
and $\boldsymbol{v}_{\textrm{A},m}^{\textrm{pri,\ensuremath{\left(t\right)}}}$
are extrinsic messages passed from Module B. According to the LMMSE
estimation, the posterior distribution of $\boldsymbol{\ddot{u}}_{m}^{\left(t\right)}$
is a Gaussian distribution with the posterior mean and covariance
given by
\begin{align}
\mathbf{V}_{\textrm{A},m}^{\textrm{post},\left(t\right)} & =\left(\hat{\gamma}^{\left(t\right)}\left(\mathbf{\ddot{G}}_{m}^{\left(t\right)}\right)^{T}\mathbf{\ddot{G}}_{m}^{\left(t\right)}+\textrm{diag}\left(1/\boldsymbol{v}_{\textrm{A},m}^{\textrm{pri,\ensuremath{\left(t\right)}}}\right)\right)^{-1},\nonumber \\
\boldsymbol{u}_{\textrm{A},m}^{\textrm{post},\left(t\right)} & =\mathbf{V}_{\textrm{A},m}^{\textrm{post},\left(t\right)}\left(\frac{\boldsymbol{u}_{\textrm{A},m}^{\textrm{pri},\left(t\right)}}{\boldsymbol{v}_{\textrm{A},m}^{\textrm{pri,\ensuremath{\left(t\right)}}}}+\hat{\gamma}^{\left(t\right)}\left(\mathbf{\ddot{G}}_{m}^{\left(t\right)}\right)^{T}\boldsymbol{\ddot{y}}_{m}^{\left(t\right)}\right).\label{eq:A_post}
\end{align}
Then, Module A outputs extrinsic messages to Module B as 
\begin{align}
\boldsymbol{u}_{\textrm{B},m}^{\textrm{pri},\left(t\right)} & =\boldsymbol{v}_{\textrm{B},m}^{\textrm{pri,\ensuremath{\left(t\right)}}}\left(\boldsymbol{u}_{\textrm{A},m}^{\textrm{post},\left(t\right)}/\boldsymbol{v}_{\textrm{A},m}^{\textrm{post},\left(t\right)}-\boldsymbol{u}_{\textrm{A},m}^{\textrm{pri},\left(t\right)}/\boldsymbol{v}_{\textrm{A},m}^{\textrm{pri,\ensuremath{\left(t\right)}}}\right),\nonumber \\
\boldsymbol{v}_{\textrm{B},m}^{\textrm{pri,\ensuremath{\left(t\right)}}} & =1/\left(1/\boldsymbol{v}_{\textrm{A},m}^{\textrm{post},\left(t\right)}-1/\boldsymbol{v}_{\textrm{A},m}^{\textrm{pri,\ensuremath{\left(t\right)}}}\right),\label{eq:A_ext}
\end{align}
where $\boldsymbol{v}_{\textrm{A},m}^{\textrm{post},\left(t\right)}=\textrm{diag}\left(\mathbf{V}_{\textrm{A},m}^{\textrm{post},\left(t\right)}\right)$.
The computational complexity of Module A is dominated by the matrix
inverse in the calculation of $\mathbf{V}_{\textrm{A},m}^{\textrm{post},\left(t\right)}$.
Fortunately, the dimension of $\mathbf{V}_{\textrm{A},m}^{\textrm{post},\left(t\right)}$
is very small thanks to the polar-delay domain filtering ($\left|\mathbf{\Omega}^{\left(t\right)}\right|$
is comparable to the number of channel paths). Besides, $M$ LMMSE
estimators can work in a parallel fashion. Therefore, Module A is
highly efficient with a relatively low computational overhead.

\subsubsection{Structured Sparse Inference in Module B}

\begin{figure}[t]
\centering{}\includegraphics[width=60mm]{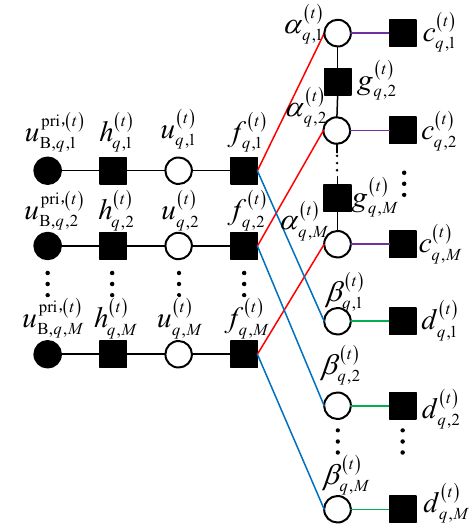}\caption{\label{fig:Factor_graph}The structure of the sub-graph $\mathcal{G}_{q}^{\left(t\right)},q\in\mathbf{\Omega}^{\left(t\right)}$.}
\end{figure}

In Module B, the extrinsic message $\boldsymbol{u}_{\textrm{B},m}^{\textrm{pri},\left(t\right)}$
is viewed as an AWGN observation of $\boldsymbol{\ddot{u}}_{m}^{\left(t\right)}$
\cite{LiuAn_CE_Turbo_CS}:
\begin{equation}
u_{\textrm{B},q,m}^{\textrm{pri},\left(t\right)}=u_{q,m}^{\left(t\right)}+n_{q,m}^{\left(t\right)},q\in\mathbf{\Omega}^{\left(t\right)},\label{eq:u_B,qi,m_pri(t}
\end{equation}
where $n_{q,m}^{\left(t\right)}\sim\mathcal{N}\left(n_{q,m}^{\left(t\right)};0,v_{\textrm{B},q,m}^{\textrm{pri,\ensuremath{\left(t\right)}}}\right)$
is the equivalent noise, $u_{\textrm{B},q,m}^{\textrm{pri},\left(t\right)}$
and $v_{\textrm{B},q,m}^{\textrm{pri},\left(t\right)}$ represent
the element of $\boldsymbol{u}_{\textrm{B},m}^{\textrm{pri},\left(t\right)}$
and $\boldsymbol{v}_{\textrm{B},m}^{\textrm{pri},\left(t\right)}$,
respectively. Then, the joint distribution associated with Module
B is given by
\begin{align}
p_{\textrm{B}}= & \prod_{q\in\mathbf{\Omega}^{\left(t\right)}}^{Q}p\left(\boldsymbol{u}_{q}^{\left(t\right)}\mid\boldsymbol{\alpha}_{q}^{\left(t\right)},\boldsymbol{\beta}_{q}^{\left(t\right)}\right)\hat{p}\left(\boldsymbol{\alpha}_{q}^{\left(t\right)}\right)\hat{p}\left(\boldsymbol{\beta}_{q}^{\left(t\right)}\right)\nonumber \\
 & \times\prod_{m=1}^{M}p\left(u_{\textrm{B},q,m}^{\textrm{pri},\left(t\right)}\mid u_{q,m}^{\left(t\right)}\right).\label{eq:joint_Moulde_B}
\end{align}
The factor graph of the joint distribution is shown in Fig. \ref{fig:Factor_graph},
where the associated factor nodes are listed in Table \ref{tab:factor_node}.
According to the structure of the joint distribution in (\ref{eq:joint_Moulde_B}),
the whole factor graph consists of $\left|\mathbf{\Omega}^{\left(t\right)}\right|$
independent sub-graphs with the same internal structure, denoted by
$\mathcal{G}_{q}^{\left(t\right)}$ for $q\in\mathbf{\Omega}^{\left(t\right)}$.
We use the sum-product rule \cite{Ksch_sum-product} to perform message
passing over each sub-graph $\mathcal{G}_{q}^{\left(t\right)}$.
\begin{table}[t]
\centering{}\caption{\label{tab:factor_node}Factors, distributions, and functional forms
in Fig. \ref{fig:Factor_graph}.}
\begin{tabular}{|c|c|c|}
\hline 
\multirow{1}{*}{Factor} & \multirow{1}{*}{Distribution} & Functional form\tabularnewline
\hline 
\hline 
$h_{q,m}^{\left(t\right)}$ & $p\left(u_{\textrm{B},q,m}^{\textrm{pri},\left(t\right)}\mid u_{q,m}^{\left(t\right)}\right)$ & $\mathcal{N}\left(u_{\textrm{B},q,m}^{\textrm{pri},\left(t\right)};u_{q,m}^{\left(t\right)},v_{\textrm{B},q,m}^{\textrm{pri,\ensuremath{\left(t\right)}}}\right)$\tabularnewline
\hline 
$f_{q,m}^{\left(t\right)}$ & $p\left(u_{q,m}^{\left(t\right)}\mid\alpha_{q,m}^{\left(t\right)},\beta_{q,m}^{\left(t\right)}\right)$ & $\delta\left(u_{q,m}^{\left(t\right)}-\alpha_{q,m}^{\left(t\right)}\beta_{q,m}^{\left(t\right)}\right)$\tabularnewline
\hline 
$c_{q,m}^{\left(t\right)}$ & $\hat{p}\left(\alpha_{q,m}^{\left(t\right)}\right)$ & given in (\ref{eq:p^(alpha_q,m^t)})\tabularnewline
\hline 
$g_{q,m}^{\left(t\right)}$ & $p\left(\alpha_{q,m}^{\left(t\right)}\mid\alpha_{q,m-1}^{\left(t\right)}\right)$ & given in (\ref{eq:p01_S}) and (\ref{eq:p10_S})\tabularnewline
\hline 
$d_{q,m}^{\left(t\right)}$ & $\hat{p}\left(\beta_{q,m}^{\left(t\right)}\right)$ & given in (\ref{eq:p^(beta_q,m^(t))})\tabularnewline
\hline 
\end{tabular}
\end{table}

Specifically, we first calculate messages over $u_{q,m}^{\left(t\right)}/\beta_{q,m}^{\left(t\right)}\rightarrow f_{q,m}^{\left(t\right)}\rightarrow\alpha_{q,m}^{\left(t\right)}$.
Then, forward-backward message passing is performed over the Markov
chain $\alpha_{q,1}^{\left(t\right)}\rightarrow\alpha_{q,2}^{\left(t\right)}\rightarrow\ldots\rightarrow\alpha_{q,M}^{\left(t\right)}$.
Finally, we perform message passing over $\alpha_{q,m}^{\left(t\right)}\rightarrow f_{q,m}^{\left(t\right)}\rightarrow u_{q,m}^{\left(t\right)}/\beta_{q,m}^{\left(t\right)}$.
The marginal posterior distribution of $u_{q,m}^{\left(t\right)}$,
$\alpha_{q,m}^{\left(t\right)}$ and $\beta_{q,m}^{\left(t\right)}$
can be computed by
\begin{align}
q\left(u_{q,m}^{\left(t\right)}\right) & \propto\upsilon_{f_{q,m}^{\left(t\right)}\rightarrow u_{q,m}^{\left(t\right)}}\left(u_{q,m}^{\left(t\right)}\right)\upsilon_{u_{q,m}^{\left(t\right)}\rightarrow f_{q,m}^{\left(t\right)}}\left(u_{q,m}^{\left(t\right)}\right),\label{eq:p(u_q1,m_(t)|u_B,q1,m_pri,(t))}\\
q\left(\alpha_{q,m}^{\left(t\right)}\right) & \propto\upsilon_{f_{q,m}^{\left(t\right)}\rightarrow\alpha_{q,m}^{\left(t\right)}}\left(\alpha_{q,m}^{\left(t\right)}\right)\upsilon_{\alpha_{q,m}^{\left(t\right)}\rightarrow f_{q,m}^{\left(t\right)}}\left(\alpha_{q,m}^{\left(t\right)}\right),\label{eq:p(alpha_q,m)_t}\\
q\left(\beta_{q,m}^{\left(t\right)}\right) & \propto\upsilon_{f_{q,m}^{\left(t\right)}\rightarrow\beta_{q,m}^{\left(t\right)}}\left(\beta_{q,m}^{\left(t\right)}\right)\upsilon_{\beta_{q,m}^{\left(t\right)}\rightarrow f_{q,m}^{\left(t\right)}}\left(\beta_{q,m}^{\left(t\right)}\right),\label{eq:q(beta_q,m^(t))}
\end{align}
where $\upsilon_{f_{q,m}^{\left(t\right)}\rightarrow u_{q,m}^{\left(t\right)}}$
means the message passed from factor node $f_{q,m}^{\left(t\right)}$
to variable node $u_{q,m}^{\left(t\right)}$. And the posterior mean
$u_{\textrm{B},q,m}^{\textrm{post},\left(t\right)}$ and variance
$v_{\textrm{B},q,m}^{\textrm{post},\left(t\right)}$ of $u_{q,m}^{\left(t\right)}$
can be obtained based on (\ref{eq:p(u_q1,m_(t)|u_B,q1,m_pri,(t))}).
Denote $\boldsymbol{u}_{\textrm{B},m}^{\textrm{post},\left(t\right)}\triangleq\left[u_{\textrm{B},q,m}^{\textrm{post},\left(t\right)}\right]_{q\in\mathbf{\Omega}^{\left(t\right)}}$
and $\boldsymbol{v}_{\textrm{B},m}^{\textrm{post},\left(t\right)}\triangleq\left[v_{\textrm{B},q,m}^{\textrm{post},\left(t\right)}\right]_{q\in\mathbf{\Omega}^{\left(t\right)}}$,
the extrinsic message passed from Module A to Module B is given by
\begin{align}
\boldsymbol{u}_{\textrm{A},m}^{\textrm{pri},\left(t\right)} & =\boldsymbol{v}_{\textrm{A},m}^{\textrm{pri,\ensuremath{\left(t\right)}}}\left(\boldsymbol{u}_{\textrm{B},m}^{\textrm{post},\left(t\right)}/\boldsymbol{v}_{\textrm{B},m}^{\textrm{post},\left(t\right)}-\boldsymbol{u}_{\textrm{B},m}^{\textrm{pri},\left(t\right)}/\boldsymbol{v}_{\textrm{B},m}^{\textrm{pri,\ensuremath{\left(t\right)}}}\right),\nonumber \\
\boldsymbol{v}_{\textrm{A},m}^{\textrm{pri,\ensuremath{\left(t\right)}}} & =1/\left(1/\boldsymbol{v}_{\textrm{B},m}^{\textrm{post},\left(t\right)}-1/\boldsymbol{v}_{\textrm{B},m}^{\textrm{pri,\ensuremath{\left(t\right)}}}\right).\label{eq:B_ext}
\end{align}
After convergence of the algorithm, the MAP estimate of $u_{q,m}^{\left(t\right)}\in\left[0,+\infty\right),q\in\mathbf{\Omega}^{\left(t\right)},\forall m$
is updated as
\begin{equation}
\hat{u}_{q,m}^{\left(t\right)}=\begin{cases}
u_{\textrm{A},q,m}^{\textrm{post},\left(t\right)}, & u_{\textrm{A},q,m}^{\textrm{post},\left(t\right)}\geq0\\
0, & u_{\textrm{A},q,m}^{\textrm{post},\left(t\right)}<0
\end{cases}.\label{eq:Uq1_hat_(t)}
\end{equation}

\subsection{Grid Update Module}

Given $\hat{\boldsymbol{x}}^{\left(t\right)}$, $\hat{\gamma}^{\left(t\right)}$,
and $\hat{\mathbf{U}}^{\left(t\right)}$, the log-likelihood function
is given by
\begin{align}
\mathcal{L}\left(\mathbf{\Xi}^{\left(t\right)}\right) & =\ln p\left(\tilde{\boldsymbol{y}}^{\left(t\right)}\mid\hat{\boldsymbol{x}}^{\left(t\right)},\hat{\mathbf{U}}^{\left(t\right)},\hat{\gamma}^{\left(t\right)};\mathbf{\Xi}^{\left(t\right)}\right)\\
 & =-\hat{\gamma}^{\left(t\right)}\left\Vert \tilde{\boldsymbol{y}}^{\left(t\right)}-\mathbf{F}\left(\mathbf{\Xi}^{\left(t\right)},\hat{\mathbf{U}}^{\left(t\right)}\right)\hat{\boldsymbol{x}}^{\left(t\right)}\right\Vert ^{2}+C,\nonumber 
\end{align}
where $C$ is a constant. It is intractable to directly find the global
optimal solution that maximizes $\mathcal{L}\left(\mathbf{\Xi}^{\left(t\right)}\right)$
since it is non-concave w.r.t. $\mathbf{\Xi}^{\left(t\right)}$. And
thus, we update grid parameters via gradient ascent. $\mathbf{\Xi}^{\left(t\right)}$
can be divided into $J=3$ blocks with $\mathbf{\Xi}_{1}^{\left(t\right)}=\boldsymbol{\vartheta}^{\left(t\right)}$,
$\mathbf{\Xi}_{2}^{\left(t\right)}=\frac{1}{\boldsymbol{r}^{\left(t\right)}}$,
and $\mathbf{\Xi}_{3}^{\left(t\right)}=\boldsymbol{\boldsymbol{\tau}}^{\left(t\right)}$
according to different physical meanings. Starting from the initial
point $\mathbf{\Xi}_{j}^{\left(t\right)\left(0\right)}=\hat{\mathbf{\Xi}}^{\left(t-1\right)}$,
where $\hat{\mathbf{\Xi}}^{\left(t-1\right)}$ is the estimated grid
passed from time $t-1$, in the $i\textrm{-th}$ iteration, each block
is updated as
\begin{equation}
\mathbf{\Xi}_{j}^{\left(t\right)\left(i\right)}=\mathbf{\Xi}_{j}^{\left(t\right)\left(i-1\right)}+\epsilon_{j}^{\left(i\right)}\left.\tfrac{\partial\mathcal{L}\left(\mathbf{\Xi}_{j}^{\left(t\right)},\mathbf{\Xi}_{-j}^{\left(t\right)\left(i\right)}\right)}{\partial\mathbf{\Xi}_{j}^{\left(t\right)}}\right|{}_{\mathbf{\Xi}_{j}^{\left(t\right)}=\mathbf{\Xi}_{j}^{\left(t\right)\left(i-1\right)}},\label{eq:gradient_update}
\end{equation}
where $\mathbf{\Xi}_{-j}^{\left(t\right)\left(i\right)}\triangleq\left(\mathbf{\Xi}_{1}^{\left(t\right)\left(i\right)},\ldots,\mathbf{\Xi}_{j-1}^{\left(t\right)\left(i\right)},\mathbf{\Xi}_{j+1}^{\left(t\right)\left(i-1\right)},\ldots,\mathbf{\Xi}_{J}^{\left(t\right)\left(i-1\right)}\right)$,
and $\epsilon_{j}^{\left(i\right)}$ is the step size determined by
the Armijo rule.

\subsection{Temporal Correlated Module}

In this module, we calculate $\hat{p}\left(\boldsymbol{s}^{\left(t+1\right)}\right)$,
$\hat{p}\left(\boldsymbol{\alpha}_{q}^{\left(t+1\right)}\right)$,
and $\hat{p}\left(\boldsymbol{\beta}_{q}^{\left(t+1\right)}\right)$
as the prior for time $t+1$.

\subsubsection{Update of $\hat{p}\left(\boldsymbol{s}^{\left(t+1\right)}\right)$}

Given the posterior distribution $q\left(s_{q}^{\left(t\right)}\right),\forall q$
output by the IF-VBI, $\hat{p}\left(s_{q}^{\left(t+1\right)}\right)$
is update as $\hat{p}\left(s_{q}^{\left(t+1\right)}=1\right)=\lambda_{q}^{\left(t+1\right)},$where
\begin{equation}
\lambda_{q}^{\left(t+1\right)}=p_{11}^{x}q\left(s_{q}^{\left(t\right)}=1\right)+p_{01}^{x}q\left(s_{q}^{\left(t\right)}=0\right).\label{eq:lambda_q(t+1)}
\end{equation}

\subsubsection{Update of $\hat{p}\left(\boldsymbol{\alpha}_{q}^{\left(t+1\right)}\right)$}

Given $q\left(\alpha_{q,m}^{\left(t\right)}\right)$ in (\ref{eq:p(alpha_q,m)_t}),
$\hat{p}\left(\alpha_{q,m}^{\left(t+1\right)}\right),\forall m$ is
calculated as
\begin{equation}
\hat{p}\left(\alpha_{q,m}^{\left(t+1\right)}=1\right)=\begin{cases}
\pi_{q,m}^{\textrm{pri},\left(t+1\right)}, & q\in\mathbf{\Omega}^{\left(t\right)}\\
\kappa, & q\notin\mathbf{\Omega}^{\left(t\right)}
\end{cases},\label{eq:p^(alpha_q,m^t)}
\end{equation}
where
\begin{equation}
\pi_{q,m}^{\textrm{pri},\left(t+1\right)}=p_{11}^{T}q\left(\alpha_{q,m}^{\left(t\right)}=1\right)+p_{01}^{T}q\left(\alpha_{q,m}^{\left(t\right)}=0\right).\label{eq:pi_q1,m_pri,(t+1)}
\end{equation}

\subsubsection{Update of $\hat{p}\left(\boldsymbol{\beta}_{q}^{\left(t+1\right)}\right)$}

Denote $q\left(\beta_{q,m}^{\left(t\right)}\right)$ in (\ref{eq:q(beta_q,m^(t))})
as $\mathcal{N}\left(\beta_{q,m}^{\left(t\right)};\mu_{q,m}^{\textrm{post},\left(t\right)},\nu_{q,m}^{\textrm{post},\left(t\right)}\right)$,
we update $\hat{p}\left(\beta_{q,m}^{\left(t+1\right)}\right)$ as
\begin{equation}
\hat{p}\left(\beta_{q,m}^{\left(t+1\right)}\right)=\begin{cases}
\mathcal{N}\left(\beta_{q,m}^{\left(t+1\right)};\mu_{q,m}^{\left(t+1\right)},\nu_{q,m}^{\left(t+1\right)}\right), & q\in\mathbf{\Omega}^{\left(t\right)}\\
\mathcal{N}\left(\beta_{q,m}^{\left(t+1\right)};\zeta,\frac{\varepsilon\sigma^{2}}{2-\varepsilon}\right), & q\notin\mathbf{\Omega}^{\left(t\right)}
\end{cases},\label{eq:p^(beta_q,m^(t))}
\end{equation}
with $\mu_{q,m}^{\left(t+1\right)}$ and $\nu_{q,m}^{\left(t+1\right)}$
given by
\begin{align}
\mu_{q,m}^{\left(t+1\right)} & =\left(1-\varepsilon\right)\mu_{q,m}^{\textrm{post},\left(t\right)}+\varepsilon\zeta,\nonumber \\
\nu_{q,m}^{\left(t+1\right)} & =\left(1-\varepsilon\right)^{2}\nu_{q,m}^{\textrm{post},\left(t\right)}+\varepsilon^{2}\sigma^{2}.\label{eq:prior of beta_q1,m_(t+1)}
\end{align}

Moreover, the estimated grid $\hat{\mathbf{\Xi}}^{\left(t\right)}$
is used as the initial value of $\mathbf{\Xi}^{\left(t+1\right)}$
to accelerate the convergence speed of signal processing at time $t+1$.

\subsection{Complexity Analysis}

The DA-MAP algorithm is summarized in Algorithm \ref{DA_MAP}. The
IF-VBI estimator avoids the matrix inverse, and its complexity is
reduced to $\mathcal{O}\left(MQ\right)$ per iteration \cite{Xu_alternating_MAP}.
The complexity of the Turbo-CS is dominated by the small-scale matrix
inverse in (\ref{eq:A_post}), whose complexity is $\mathcal{O}\left(M\left|\mathbf{\Omega}^{\left(t\right)}\right|^{3}\right)$
per iteration. Note that the message passing in Module B of the Turbo-CS
and the temporal correlated module is in linear complexity, which
is almost negligible. The complexity of the gradient calculation in
(\ref{eq:gradient_update}) is $\mathcal{O}\left(Q^{2}\right)$. Let
$I_{1}$ and $I_{2}$ represent the inner iteration number of IF-VBI
and Turbo-CS, respectively, and let $I$ denote the outer iteration
number of DA-MAP. Then, the overall complexity of the DA-MAP is $\mathcal{O}\left(I\left(I_{1}MQ+I_{2}M\left|\mathbf{\Omega}^{\left(t\right)}\right|^{3}+Q^{2}\right)\right)$.
\begin{algorithm}[tbh]
\begin{singlespace}
{\small\caption{\label{DA_MAP}The DA-MAP algorithm}
}{\small\par}

\textbf{Input:} $\tilde{\boldsymbol{y}}^{\left(1:T\right)}$, initial
grid $\hat{\mathbf{\Xi}}^{\left(0\right)}=\boldsymbol{\bar{\Xi}}$,
and initial VR matrix $\hat{\mathbf{U}}^{\left(0\right)}=\mathbf{1}_{M\times Q}$.

\textbf{Output:} $\hat{\boldsymbol{x}}^{\left(1:T\right)},\mathbf{\hat{U}}^{\left(1:T\right)},\hat{\mathbf{\Xi}}^{\left(1:T\right)}$.

\begin{algorithmic}[1]

\FOR{${\color{blue}{\color{black}t=1,\cdots,T}}$}

\STATE Initialize: $\hat{\mathbf{U}}^{\left(t\right)}=\hat{\mathbf{U}}^{\left(t-1\right)}$,
$\hat{\mathbf{\Xi}}^{\left(t\right)}=\hat{\mathbf{\Xi}}^{\left(t-1\right)}$.

\FOR{${\color{blue}{\color{black}i=1,\cdots,I}}$}

\STATE\textbf{\% Channel estimation module: IF-VBI}

\STATE Update $\hat{\boldsymbol{x}}$ and $\hat{\gamma}$ based on
the posterior distributions.

\STATE\textbf{\% VR detection module: Turbo-CS}

\FOR{${\color{blue}{\color{black}i_{2}=1,\cdots,I_{2}}}$}

\STATE\textbf{\% Module A: LMMSE}

\STATE Update $\boldsymbol{u}_{\textrm{A},m}^{\textrm{post},\left(t\right)}$
and $\boldsymbol{v}_{\textrm{A},m}^{\textrm{post},\left(t\right)}$,
using (\ref{eq:A_post}).

\STATE Calculate the extrinsic messages from Module A to B, using
(\ref{eq:A_ext}).

\STATE\textbf{\% Module B: Structured sparse inference}

\STATE Perform message passing over each sub-graph $\mathcal{G}_{q}^{\left(t\right)}$.

\STATE Calculate $q\left(\boldsymbol{u}_{q}^{\left(t\right)}\right)$,
$q\left(\boldsymbol{\alpha}_{q}^{\left(t\right)}\right)$, and $q\left(\boldsymbol{\beta}_{q}^{\left(t\right)}\right)$,
using (\ref{eq:p(u_q1,m_(t)|u_B,q1,m_pri,(t))}) - (\ref{eq:q(beta_q,m^(t))}).

\STATE Calculate the extrinsic messages from Module B to A, using
(\ref{eq:B_ext}).

\ENDFOR

\STATE Update $\hat{\mathbf{U}}^{\left(t\right)}$, using (\ref{eq:Uq1_hat_(t)}).

\STATE\textbf{\% Grid update module}

\STATE Update $\hat{\boldsymbol{\Xi}}^{\left(t\right)}$ via gradient
ascent, using (\ref{eq:gradient_update}).

\ENDFOR

\STATE\textbf{\% Temporal correlated module}

\STATE Calculate $\hat{p}\left(\boldsymbol{s}^{\left(t+1\right)}\right)$,
$\hat{p}\left(\boldsymbol{\alpha}_{q}^{\left(t+1\right)}\right)$,
and $\hat{p}\left(\boldsymbol{\beta}_{q}^{\left(t+1\right)}\right)$,
using (\ref{eq:lambda_q(t+1)}) - (\ref{eq:prior of beta_q1,m_(t+1)}).

\ENDFOR

\end{algorithmic}
\end{singlespace}
\end{algorithm}

\section{Simulation Results}

In this section, we evaluate the channel tracking performance of the
proposed DA-MAP through comprehensive simulations. Some benchmarks
and the proposed method are summarized below.
\begin{itemize}
\item \textbf{OMP \cite{HanYu_VR,Chen_NS_CE}:} The off-grid OMP estimates
spatial stationary sub-channels of each antenna independently in the
delay domain.
\item \textbf{Turbo-OAMP \cite{Vincent_sparse_modeling}:} The off-grid
Turbo-OAMP is employed to estimate the antenna-delay domain channels. 
\item \textbf{Two-stage message passing (MP) \cite{Tang_sparse_modeling}:}
We extend the two-stage scheme in \cite{Tang_sparse_modeling} from
the single-path scenario to the multi-path scenario. In stage 1, the
Turbo-OAMP estimates continuous VRs based on a coarse estimate of
the channel vector; in stage 2, the off-grid OMP recovers the polar-delay
domain channel vector precisely. 
\item \textbf{Alternating MAP \cite{Xu_alternating_MAP}: }This is the work
more directly relevant to the proposed DA-MAP. The main drawback is
that the structured EP can only recover binary VRs.
\item \textbf{DA-MAP (i.i.d.): }It is the proposed DA-MAP with i.i.d. Bernoulli
priors for channel supports and VRs. In this case, the temporal correlation
of channel supports and structured sparsity of VRs cannot be exploited. 
\item \textbf{DA-MAP (Markov):} It is the proposed DA-MAP with structured
Markov priors.
\end{itemize}

The parameters of the broadband XL-MIMO system are set as follows:
the number of ULA antennas and RF chains at the BS is $M=256$ and
$N_{\textrm{RF}}=64$, respectively; the center frequency is set to
$f_{c}=28\ \textrm{GHz}$, and the Rayleigh distance is $\frac{2D^{2}}{\lambda_{c}}=348\ \textrm{m}$;
the subcarrier interval is $f_{0}=120\ \textrm{kHz}$ and the number
of subcarriers assigned to each user is $N=64$; the number of frames
is $T=20$ with each frame containing a pilot sequence of length $P=M_{\textrm{sub}}=4$.
We use a 3GPP-like channel simulation toolbox called NF-SnS \cite{3GPP_NF_SnS}
to generate XL-MIMO channels with power change. The user speed is
set to $3\ \textrm{km/h}$ and the time interval between two adjacent
frames is $50\ \textrm{ms}$. The initial number of channel paths
is $L^{\left(1\right)}=4$ and the visibility probability is $\kappa=0.5$.
We use the normalized mean square error (NMSE) to measure the channel
estimation performance. Besides, the VR estimation NMSE is the performance
metric for VR detection, which is defined as
\begin{align}
\textrm{VR\ NMSE} & \triangleq\frac{1}{T}\sum_{t=1}^{T}\frac{\sum_{l=1}^{L^{\left(t\right)}}\left\Vert \boldsymbol{u}_{l}^{\left(t\right)}-\hat{\boldsymbol{u}}_{q_{l}}^{\left(t\right)}\right\Vert ^{2}}{\sum_{l=1}^{L^{\left(t\right)}}\left\Vert \boldsymbol{u}_{l}^{\left(t\right)}\right\Vert ^{2}},
\end{align}
where $q_{l}$ is the index of the polar-delay domain grid point nearest
to scatterer $l$. The received SNR is defined as $10\log\frac{\left\Vert \boldsymbol{h}^{\left(t\right)}\right\Vert ^{2}}{\left\Vert \tilde{\boldsymbol{z}}^{\left(t\right)}\right\Vert ^{2}}$.
\begin{figure}[t]
\centering{}\includegraphics[width=70mm]{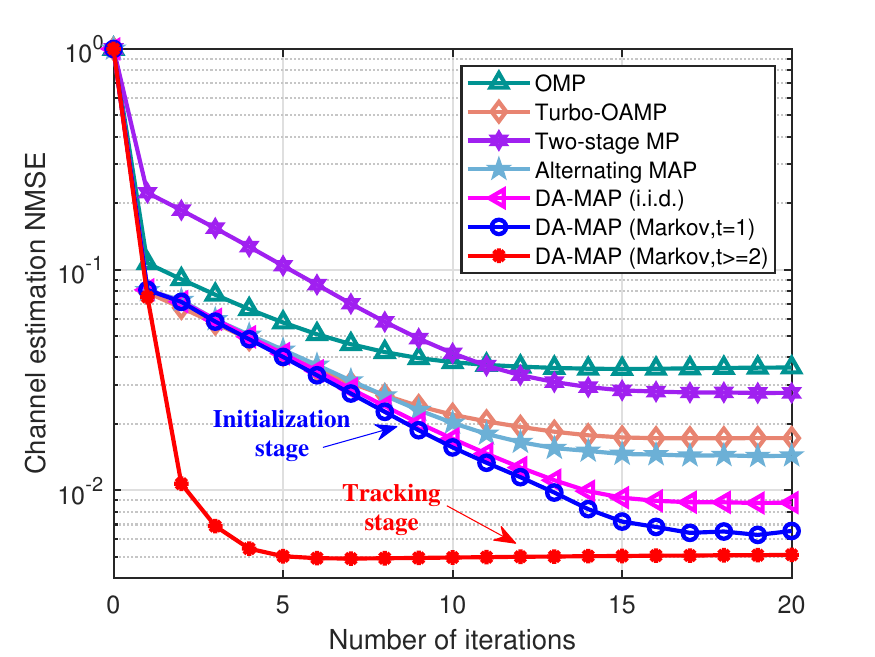}\caption{\label{fig:Convergence_NMSE}Convergence behavior: NMSE of channel
estimation with respect to the number of iterations. ($\textrm{SNR}=5\ \textrm{dB}$)}
\end{figure}

\subsection{Convergence Behavior}

The convergence behavior of different methods are shown in Fig. \ref{fig:Convergence_NMSE}.
As can be seen, the steady-state performance of the proposed method
is better than that of baselines. In the channel initialization stage
(i.e., $t=1$), there is no prior information for the first frame,
and the proposed DA-MAP converges to a good stationary point within
$15$ iterations. While in the channel tracking stage (i.e., $t\geq2$
), the convergence speed of the DA-MAP is very quick due to the strong
prior information passed from the previous frame. In this case, the
DA-MAP can converge within $5$ iterations. Besides, the steady-state
performance in the tracking stage is better than that in the initialization
stage, which reflects that the temporal correlated module in the DA-MAP
can not only speed up convergence but also enhance channel tracking
performance.
\begin{figure}[t]
\begin{centering}
\includegraphics[clip,width=70mm]{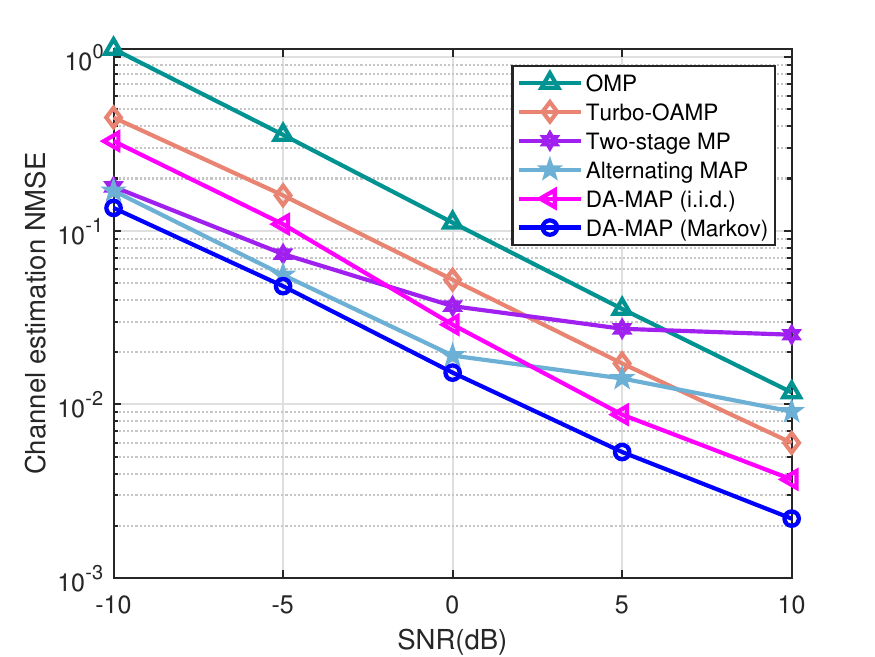}
\par\end{centering}
\centering{}\caption{\textcolor{blue}{\label{fig:NMSE_SNR}}Channel estimation NMSE versus
SNR.}
\end{figure}

\subsection{Impact of SNR}

In Fig. \ref{fig:NMSE_SNR}, we present the channel estimation NMSE
versus SNR. It is obvious that the performance of all the methods
improves as SNR increases. However, we find that the channel estimation
NMSE of the two-stage MP and alternating MAP decreases slowly in the
high SNR regions, while the channel estimation NMSE of other methods
decreases linearly with SNR. The main reason is on the VR detection
module. Specifically, the two-stage MP performs VR detection based
on a coarse estimate of channels, and thus the VR detection error
is relatively large. The alternating MAP does not consider the power
change effect among visible antennas, and the binary 0/1 VR modeling
causes a model mismatch for VRs. In the low SNR regions, the VR detection
error is small compared to the noise power. Therefore, the channel
estimation NMSE of the two baselines can still decrease linearly with
SNR. While in the high SNR regions, the noise power is small. In this
case, the VR detection error cannot be neglected and it leads to a
relatively poor estimate of channels. In contrast, the OMP and Turbo-OAMP
address the VR issue by directly estimating the spatial stationary
antenna-delay domain channels. And the proposed DA-MAP can recovery
continuous VRs accurately based on the precise VR modeling and efficient
algorithm design. Besides, the proposed DA-MAP works better than the
OMP and Turbo-OAMP since they cannot exploit the polar-domain sparsity
of the XL-MIMO channel. Moreover, the DA-MAP with structured Markov
priors achieve a significant performance gain over the DA-MAP with
i.i.d. priors, which implies that the temporal correlation of channels
and the clustered sparsity of VRs can be exploited to enhance the
performance.

\begin{figure}[t]
\begin{centering}
\includegraphics[clip,width=70mm]{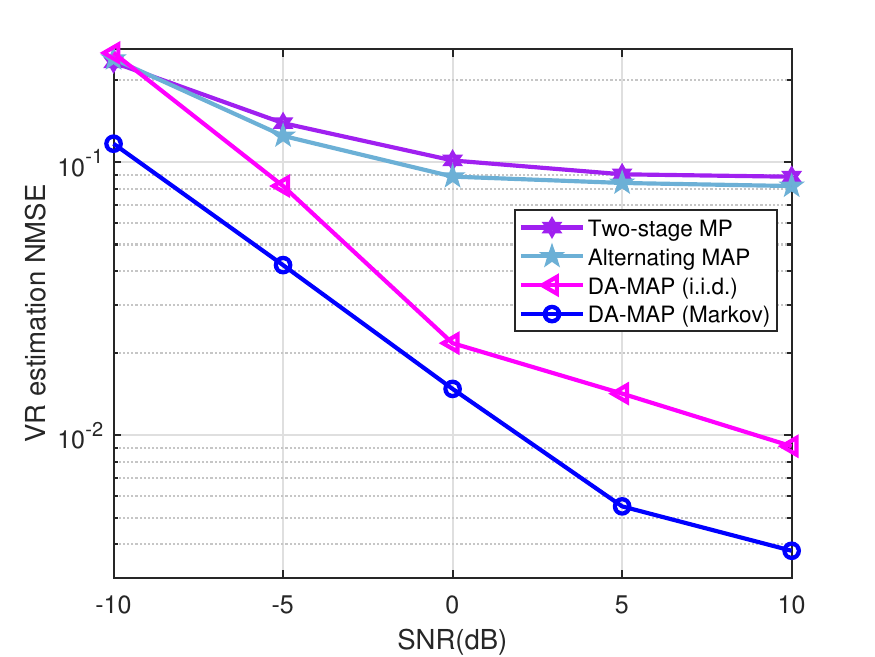}
\par\end{centering}
\caption{\label{fig:VR_NMSE_SNR}VR estimation NMSE versus SNR.}

\begin{centering}
\subfloat[SNR = $-5\ \textrm{dB}$.]{\begin{centering}
\includegraphics[width=80mm]{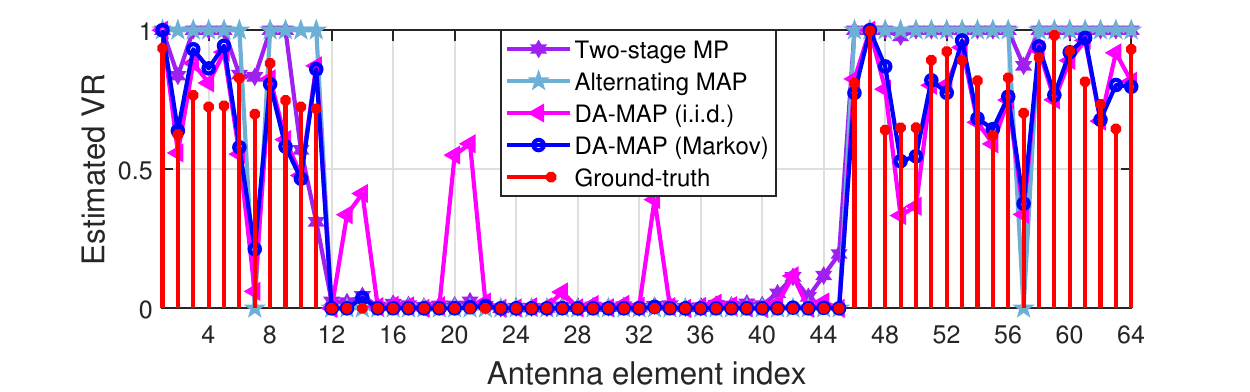}
\par\end{centering}
}
\par\end{centering}
\begin{centering}
\vspace{-3mm}
\subfloat[SNR = $5\ \textrm{dB}$.]{\begin{centering}
\includegraphics[width=80mm]{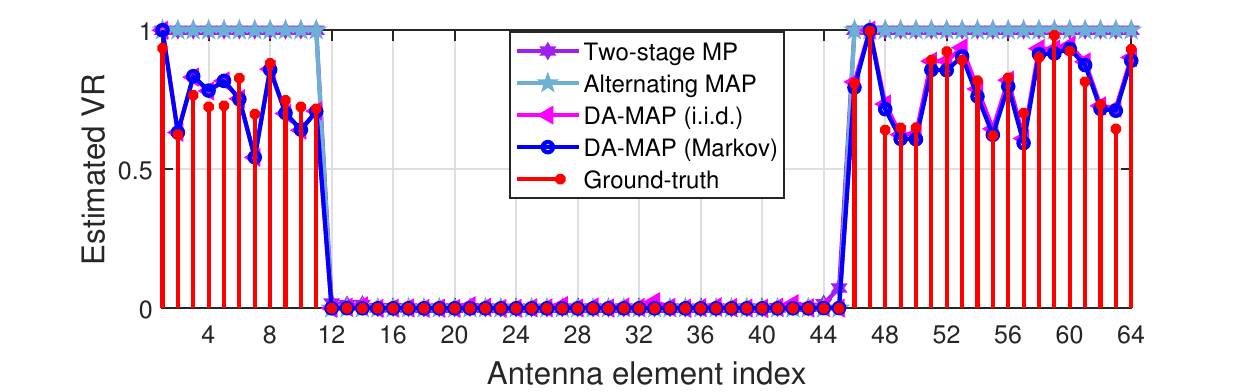}
\par\end{centering}

}
\par\end{centering}
\caption{\label{fig:VR_power_change}The power change among visible antennas
and the estimated VR of different methods.}

\end{figure}

Fig. \ref{fig:VR_NMSE_SNR} shows the VR estimation NMSE versus SNR.
As discussed previously, both the two-stage MP and alternating MAP
have a relatively larger VR estimation error than the DA-MAP. Besides,
the performance gap between the DA-MAP with Markov priors and the
same algorithm with i.i.d. priors is obvious, which means that the
proposed hierarchical 2D Markov model can capture the specific sparse
structure of VRs.

In Fig. \ref{fig:VR_power_change}, we show the estimated continuous
VR over antennas output by different methods, where both the real
VR and estimated VR are normalized within the range $\left[0,1\right]$.
It can be seen that the two-stage MP and alternating MAP can only
detect whether the antenna is visible or not, while the power of visible
antennas cannot be recovered. In contrast, the proposed DA-MAP with
Markov priors can estimate the power change among visible antennas
accurately. When $\textrm{SNR}=5\ \textrm{dB}$, the estimated VR
of the DA-MAP with Markov priors is very close to the ground truth.

\subsection{Impact of Number of Channel Paths}

\begin{figure}[t]
\centering{}\includegraphics[width=70mm]{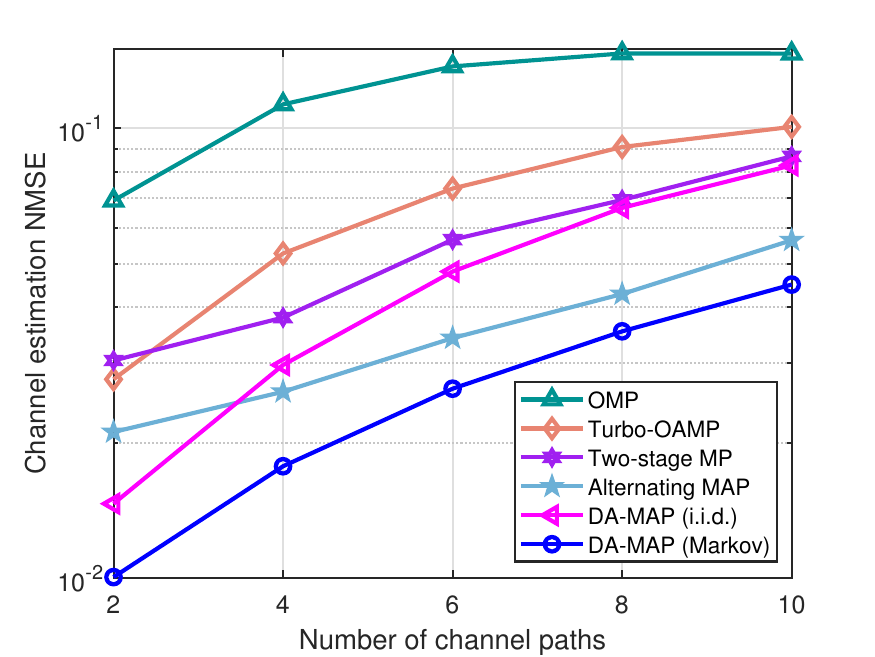}\caption{\label{fig:NMSE_nParth}Channel estimation NMSE versus the number
of channel paths. SNR is set to $0\ \textrm{dB}$.}
\end{figure}

In Fig. \ref{fig:NMSE_nParth}, we focus on how the sparsity level
of channels affect the channel estimation performance. We change the
number of channel paths from $2$ to $10$. As the number of channel
paths increases, the performance of all the methods degrades gradually.
Again, the proposed DA-MAP with Markov priors achieves the best channel
estimation performance.

\subsection{Impact of User Speed}

\begin{figure}[t]
\centering{}\includegraphics[width=70mm]{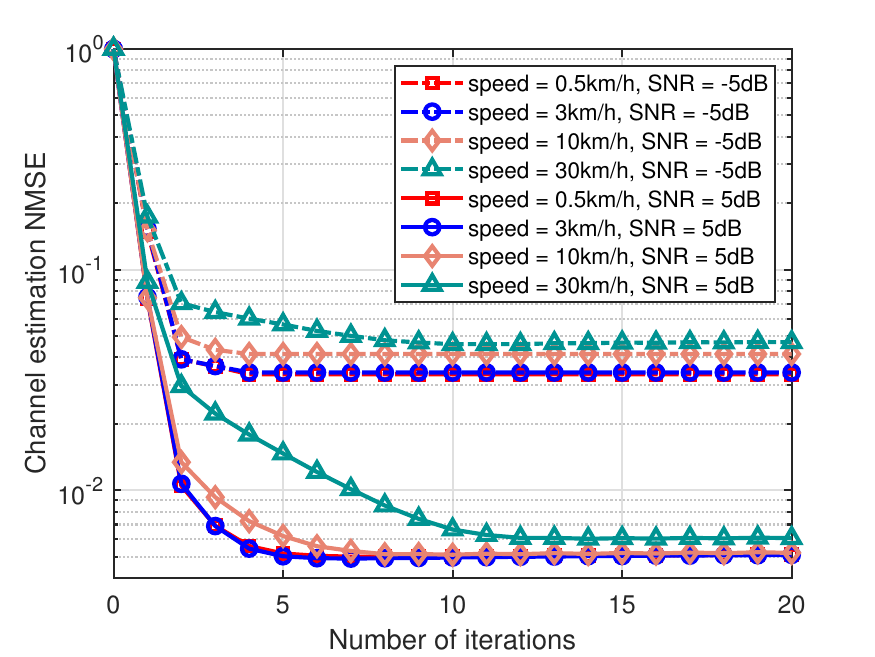}\caption{\label{fig:Channel-tracking-performance}Channel tracking performance
of the proposed method under different user speed.}
\end{figure}

In Fig. \ref{fig:Channel-tracking-performance}, we study the impact
of user speed on the channel estimation NMSE in the tracking stage.
When the user speed is high, the scattering environment changes quite
quickly. As a result, the convergence speed of the algorithm is slower
than the case of low speed. Besides, the steady-state channel tracking
performance of the low-speed scenario is better compared to the high-speed
scenario. And the performance gap between them is more obvious when
SNR is low. 

\section{Conclusions}

We propose a spatial non-stationary channel tracking scheme for XL-MIMO
systems with the power change effect. Based on the polar-delay domain
sparse representation of XL-MIMO channels, we design a three-layer
Markov prior model and a hierarchical 2D Markov model to exploit the
dynamic sparsity of channels and VRs over time, respectively. Then,
the spatial non-stationary channel tracking problem is formulated
as a bilinear measurement process. We develop a novel DA-MAP algorithm
to estimate the sparse channel vectors, detect VRs with power change,
and refine the polar-delay domain grid. Finally, simulations verified
that the proposed DA-MAP outperforms state-of-the-art baselines. 

\bibliographystyle{IEEEtran}
\bibliography{NF}

\end{document}